\documentclass[manuscript]{acmart}
\AtBeginDocument{%
  }

\setcopyright{acmlicensed}
\copyrightyear{2018}
\acmYear{2018}
\acmDOI{XXXXXXX.XXXXXXX}
\acmConference[Conference acronym 'XX]{Make sure to enter the correct
  conference title from your rights confirmation email}{June 03--05,
  2018}{Woodstock, NY}
\acmISBN{978-1-4503-XXXX-X/2018/06}

\usepackage{multirow}  
\usepackage{textcomp}  
\usepackage{booktabs}   
\usepackage{makecell}   
\usepackage{graphicx}   
\usepackage{forest}
\usepackage{wrapfig}
\usepackage{tcolorbox}

\newtcolorbox{academicbox}[1][]{
  colback=gray!10!white,   
  colframe=black!70,       
  fonttitle=\bfseries,     
  title={#1},              
  arc=1mm,                 
  boxrule=0.5pt,           
  left=6pt, right=6pt, top=4pt, bottom=4pt, 
}

\newtcolorbox{tipbox}[1][Tip]{
  colback=blue!5!white,    
  colframe=blue!80!black,  
  fonttitle=\bfseries,
  coltitle=white,          
  title={#1},
  arc=2mm,                 
  boxrule=0.8pt,           
  shadow={2mm}{-1mm}{0mm}{black!20} 
}

\newcommand\lzh[1]{{\color{black}#1}}
\newcommand\llm[1]{{\color{black}#1}}
\newcommand\rw[1]{{\color{black}#1}}

\begin{document}

\title{A Survey on Split Learning for LLM Fine-Tuning: Models, Systems, and Privacy Optimizations}


\author{Zihan Liu}
\affiliation{%
  \institution{Zhejiang University}
  \city{Ningbo}
  \country{China}}
\email{zihanliu24@zju.edu.cn}

\author{Yizhen Wang}
\affiliation{%
  \institution{Zhejiang University}
  \city{Ningbo}
  \country{China}}
\email{wyzwalker@zju.edu.cn}

\author{Rui Wang}
\affiliation{%
  \institution{Zhejiang University}
  \city{Ningbo}
  \country{China}}
\email{rwang21@zju.edu.cn}

\author{Xiu Tang}
\affiliation{%
  \institution{Zhejiang University}
  \city{Ningbo}
  \country{China}}
\email{tangxiu@zju.edu.cn}

\author{Sai Wu}
\affiliation{%
  \institution{Zhejiang University}
  \city{Ningbo}
  \country{China}}
\email{wusai@zju.edu.cn}

\authorsaddresses{%
  Authors' Contact Information: 
  Zihan Liu, Zhejiang University, Ningbo, China, zihanliu24@zju.edu.cn; 
  Yizhen Wang, Zhejiang University, Ningbo, China, wyzwalker@zju.edu.cn; 
  Rui Wang (corresponding author), Zhejiang University, Ningbo, China, rwang21@zju.edu.cn; 
  Xiu Tang, Zhejiang University, Ningbo, China, tangxiu@zju.edu.cn; 
  Sai Wu, Zhejiang University, Ningbo, China, wusai@zju.edu.cn.
}






\renewcommand{\shortauthors}{Liu et al.}

\begin{abstract}
\rw{
Fine-tuning unlocks large language models (LLMs) for specialized applications, but its high computational cost often puts it out of reach for resource-constrained organizations. While cloud platforms could provide the needed resources, data privacy concerns make sharing sensitive information with third parties risky. A promising solution is split learning for LLM fine-tuning, which divides the model between clients and a server, allowing collaborative and secure training through exchanged intermediate data, thus enabling resource-constrained participants to adapt LLMs safely.
In light of this, a growing body of literature has emerged to advance this paradigm, introducing varied model methods, system optimizations, and privacy defense-attack  techniques for split learning. To bring clarity and direction to the field, a comprehensive survey is needed to classify, compare, and critique these diverse approaches. This paper fills the gap by presenting the first extensive survey dedicated to split learning for LLM fine-tuning. We propose a unified, fine-grained training pipeline to pinpoint key operational components and conduct a systematic review of state-of-the-art work across three core dimensions: model-level optimization, system-level efficiency, and privacy preservation. Through this structured taxonomy, we establish a foundation for advancing scalable, robust, and secure collaborative LLM adaptation.

}
\end{abstract}




\maketitle

\section{Introduction}
Large Language Models (LLMs) such as GPT~\cite{brown2020language,radford2018improving},  Qwen~\cite{yang2025qwen3}, and DeepSeek~\cite{liu2024deepseek} have demonstrated exceptional capabilities in general-purpose tasks. To adapt these models to specific domains, fine-tuning has emerged as the mainstream paradigm, effectively transitioning them from {\em generalists} to specialized {\em domain experts}.
However, with the increasing scale of model parameters and the gradual depletion of high-quality public datasets, fine-tuning on private data faces the dual challenges of severe resource constraints and privacy leakage risks. Modern LLMs typically comprise tens to hundreds of billions of parameters~\cite{minaee2024llmsurvey}, rendering full fine-tuning prohibitively expensive for small and medium-sized enterprises (SMEs) with limited computational budgets. Although Parameter-Efficient Fine-Tuning (PEFT) techniques mitigate these costs by updating only a small subset of parameters, the substantial GPU memory requirements for loading and serving large models remain a significant barrier. Furthermore, fine-tuning on proprietary data raises privacy concerns regarding data outsourcing: enterprises are often hesitant to upload sensitive data to third-party cloud servers due to risks of direct exposure and potential breaches.

To address these key bottlenecks, split learning (SL)~\cite{vepakomma2018split,gupta2018distributed} has emerged as a promising distributed learning paradigm. In SL, the model is partitioned between clients and a central server: clients retain and execute only a small portion of the model (typically the front-end and/or back-end model layers), while the computationally intensive middle layers are offloaded to the server. Training proceeds via the exchange of intermediate representations (smashed data) and gradients rather than raw data, thereby significantly reducing local computational and storage burdens while preserving data privacy. Recently, combining SL with LLM fine-tuning has attracted growing research attention~\cite{lin2024splitlora, chen2024unveiling, liu2025dualguard, lin2025hsplitlora}. By applying this split-based paradigm to large language models, resource-constrained clients can participate in fine-tuning billion-parameter models while maintaining only a small fraction of parameters locally. This approach enables small and medium-sized enterprises and research institutions to collaboratively train or fine-tune private large models without exposing sensitive data to external parties, effectively addressing both resource and privacy challenges simultaneously.

Integrating split learning with LLM fine-tuning (split-based LLM-FT) has enormous potential and has spurred a wealth of research across three complementary directions. 
First, a substantial line of work targets model-level optimization challenges in multi-client federated scenarios, addressing convergence degradation caused by biased intermediate features due to data heterogeneity (Non-IID), server-client update imbalance arising from increased server-side batch sizes when aggregating smashed data, and catastrophic forgetting during sequential training or partial client participation. 
Second, another research stream focuses on system-level efficiency improvements, tackling severe communication bottlenecks from frequent transmission of massive intermediate activations and gradients, resource utilization imbalances caused by server-client coordination overhead, and the {\em straggler effect} induced by heterogeneous client computing power and network bandwidth. 
Third, a growing body of literature investigates privacy and security concerns, encompassing both attack methodologies that exploit semantically rich gradients and smashed data to reconstruct private information, as well as the distributed framework's exposed model manipulation interfaces to inject malicious behaviors, and their corresponding defense strategies.

Several surveys on split learning have already been published, as summarized in Table~\ref{tab:Comparison_of_Surveys}. 
For instance, \cite{hu2025review} categorizes SL paradigms into model-only segmentation, weight-based aggregation, and intermediate data aggregation methods, evaluating their performance under IID and non-IID scenarios. \cite{lin2024split} surveys resource-efficient frameworks and optimization strategies for alleviating computational and communication bottlenecks caused by stragglers. \cite{shabbir2025taxonomy,khan2025oops} summarize attack and defense approaches in split learning, while \cite{gu2025vflair} evaluates perturbation-based and learning-based defenses against data reconstruction and label inference attacks and their impact on downstream performance. 

\rw{
However, despite these foundational efforts, existing surveys suffer from three critical gaps in the context of LLM adaptation.
First, current surveys primarily focus on traditional, small-scale models, overlooking the distinct challenges of integrating split learning with LLMs.
From a model perspective, they assess optimization strategies in conventional tasks without verifying their efficacy for LLM fine-tuning. 
At the system level, they neglect the unique architectural demands, massive parameter scales, and performance constraints of LLMs. 
Regarding privacy, they overlook the exacerbated vulnerabilities of auto-regressive LLMs, where training labels act as shifted input sequences, making label leakage mathematically equivalent to exposing raw proprietary text.
Second, previous surveys treat split learning paradigms as macro-level architectures without establishing a unified framework for split-based LLM fine-tuning.
They lack an explicit, component-level decomposition of the computational and communication steps involved in split-based LLM fine-tuning, leaving the training pipeline ambiguous.
Third, current surveys do not systematically map multidimensional challenges and existing solutions to the specific execution stages of LLM adaptation. 
Without a unified training pipeline, there is no structured way to anchor inefficiencies, model degradations, privacy threats, and their corresponding countermeasures to precise operational components. 
This omission hinders the development of a comprehensive taxonomy needed to guide component-level design and future research directions.
}

\lzh{
}


To bridge this critical gap, this paper presents the first comprehensive review focused exclusively on the intersection of split learning and LLM fine-tuning. Our primary contributions are threefold:
\begin{table*}[t]
    \centering
    \caption{\lzh{Comparison of related surveys and research works of split learning.}}
    \renewcommand{\arraystretch}{1.1}
    \resizebox{\textwidth}{!}{
    \begin{tabular}{|c|c|c|c|>{\centering\arraybackslash}m{2.2cm}|c|c|c|}
    \hline
        \multirow{2}{*}{Paper} & \multirow{2}{*}{Year} & \multicolumn{3}{c|}{Focus Areas}  & \multirow{2}{*}{\makecell[c]{Unified Training \\ Pipeline} } & \multirow{2}{*}{\makecell[c]{Component-Level\\ Bottleneck}} & \multirow{2}{*}{LLM}  \\ 
        \cline{3-5}
        ~ & ~ & Model Performance & System Optimization & Privacy & ~ & ~ & ~ \\ \hline
        ~\cite{duan2022combined} & 2022 & \checkmark & \checkmark & \checkmark & ~ & ~ & ~ \\ \hline
        ~\cite{lin2024split} & 2024 & ~ & \checkmark & ~ & ~ & ~ & ~  \\ \hline
        ~\cite{hu2025slperf} & 2025 & \checkmark & \checkmark  & ~ & ~ & ~ & ~  \\ \hline
        ~\cite{radovivc2025towards} & 2025 & \checkmark  & \checkmark  & ~ & ~ & ~ & ~   \\ \hline
        ~\cite{hu2025review} & 2025 & \checkmark  & \checkmark  & ~ & ~ & ~ & ~   \\ \hline
        ~\cite{hukkeri2025comprehensive} & 2025 & \checkmark & \checkmark & \checkmark & ~ & ~ & ~   \\ \hline
        ~\cite{shabbir2025taxonomy} & 2025 & ~ & ~ & \checkmark & ~ & ~ & ~  \\ \hline
        ~\cite{gu2025vflair} & 2025 & ~ & ~ & \checkmark  & ~ & ~ & \checkmark   \\ \hline
        ~\cite{khan2025oops} & 2025 & ~ & ~ & \checkmark & ~ & ~ & ~   \\ \hline
        Ours & - & \checkmark & \checkmark & \checkmark & \checkmark  & \checkmark & \checkmark   \\ \hline

    \end{tabular}
    }
    \label{tab:Comparison_of_Surveys}
\end{table*}
\lzh{
\begin{itemize}
    \item A unified split-based LLM-FT pipeline: Diverging from previous coarse classifications, we abstract and decompose existing methods into a unified, highly granular training pipeline. This structural breakdown enables precise analysis of component-level bottlenecks and future optimization trajectories.
    \item multi-dimensional bottleneck analysis: We map critical vulnerabilities directly to specific execution steps, systematically exposing the root causes of convergence degradation, system inefficiencies, and privacy threats in split-based LLM-FT.
    \item component-anchored literature taxonomy: By mapping multidimensional challenges and state-of-the-art optimization methods directly onto our unified pipeline, we deliver a comprehensive taxonomy. This framework explicitly anchors system inefficiencies, model degradations, and privacy countermeasures to specific execution steps, systematically guiding future research.
\end{itemize}
}

\begin{figure*}[t]
    \centering
    \includegraphics[width=0.93\linewidth,height=330pt]{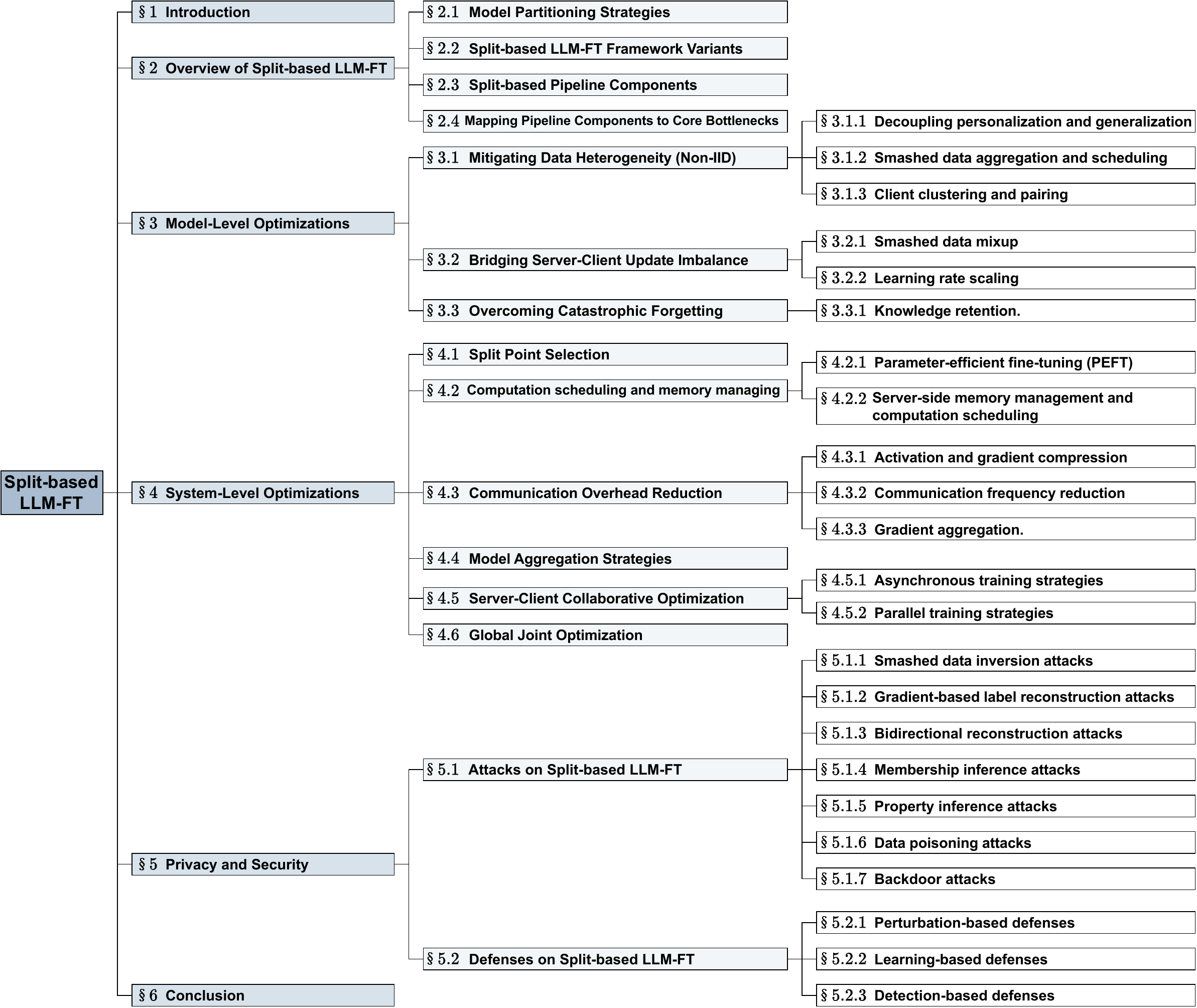}
    \vskip -5pt
    \caption{Survey framework: from a unified training pipeline to a multidimensional taxonomy of system, model, and privacy.}
    \vskip -5pt
    \label{fig:structure}
    \vskip -10pt
\end{figure*}

The rest of the article is structured as follows and the structure is in Figure~\ref{fig:structure}. 
Section~\ref{section:overview} provides a comprehensive overview of split-based LLM-FT, detailing model partitioning strategies, overarching frameworks, and our proposed unified pipeline components. \lzh{Section~\ref{section: model} systematically analyzes strategies for model-level optimization and performance enhancement. Section~\ref{section:system} explores current methodologies for system-level optimization.} Section~\ref{section: privacy} delves into the privacy and security landscape, categorizing prominent attack vectors and their corresponding defense mechanisms within split-based architectures. Finally, Section~\ref{section: conclusion} concludes this survey.

\section{Overview of Split-based LLM-FT: Model Partition, Framework Variants, and Pipeline Components}
\label{section:overview}
\subsection{Model Partitioning Strategies}

\begin{figure}[t]
\centerline{\includegraphics[width=0.95\columnwidth]{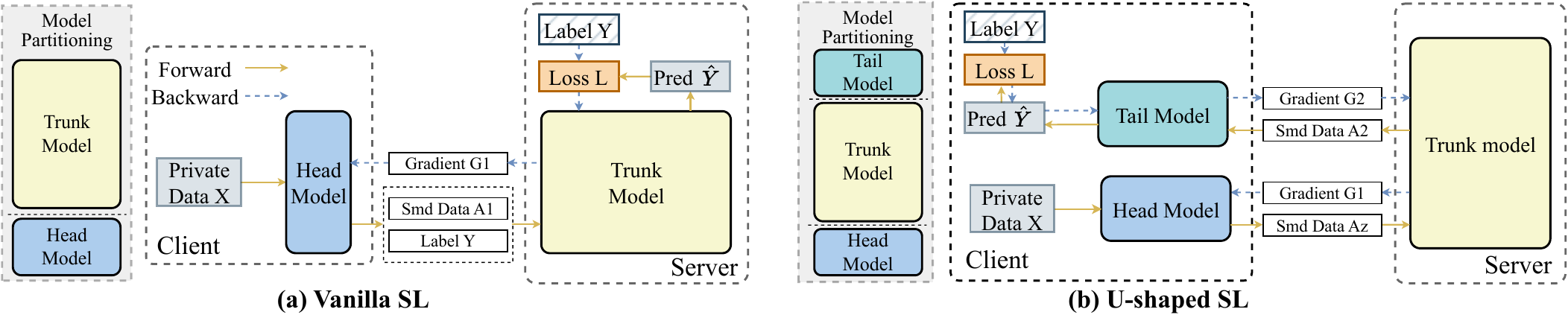}}
\vskip -5pt
\caption{Two distinct model partitioning strategies.}
\vskip -13pt
\label{fig:model_partition}
\end{figure}

Figure \ref{fig:model_partition} illustrates two primary model partitioning paradigms: Vanilla SL and U-shaped SL.

\noindent\textbf{\lzh{Two-part split (Vanilla SL)}}~\cite{vepakomma2018split,gupta2018distributed,joshi2021splitfed} partitions the model into two parts (head model-trunk model), deploying the majority of model layers on a remote server while retaining a small portion locally. During each forward pass, the client transmits forward intermediate activations and labels to the server. The server then computes the loss and performs backpropagation, returning gradients to the client. The client then performs local backpropagation. This strategy focuses on reducing computational and storage burdens on the client.  

\noindent\textbf{\lzh{Three-part split (U-shaped SL)}}~\cite{lyu2023optimal,han2023federated,chawla2024beyond} divides the model into three parts (Head Model-Trunk Model-Tail Model), keeping loss calculation and labels locally. 
\lzh{This architecture inherently enhances privacy by preventing direct label exposure. 
This protection is particularly critical in LLM supervised fine-tuning (SFT); because SFT employs teacher forcing, where labels are merely left-shifted input sequences that encode the complete raw text, localizing the loss computation is imperative to thwart server-side data reconstruction.}
However, compared to Vanilla SL, U-shaped SL strictly doubles the communication volume within a single forward-backward pass, as intermediate activations and gradients must traverse the network twice. 

\subsection{Split-based LLM-FT Framework Variants}

\begin{figure*}[t]
\centerline{\includegraphics[width=\textwidth]{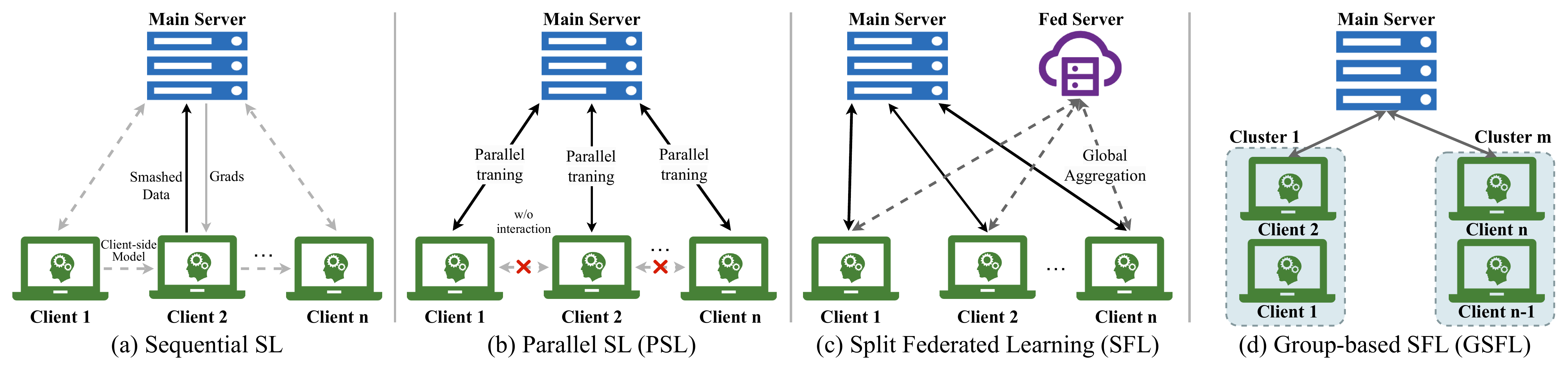}}
\vskip -5pt
\caption{Split-based LLM-FT framework variants, categorized by training coordination and aggregation mechanisms.}
\vskip -20pt
\label{fig:SL_Frameworks}
\end{figure*}

Based on different training strategies, the split learning paradigm can be categorized into several training frameworks, as illustrated in Figure~\ref{fig:SL_Frameworks}.
\lzh{Although these architectural variants were initially designed for conventional split learning tasks, their coordination mechanisms can be seamlessly adapted to collaborative LLM fine-tuning scenarios.}

\noindent\textbf{Sequential split learning (Sequential SL)~\cite{vepakomma2018split,gupta2018distributed}.} Sequential SL is the earliest proposed Split Learning framework, in which clients train sequentially with the server. After completing their local training, each client transfers its updated local model to the next client in a predefined order. In this manner, clients take turns training on their respective data partitions, forming a serialized training workflow.

\noindent\textbf{Parallel split learning (PSL)~\cite{lin2024efficient,jeon2020privacy}.} To address the inefficiency of Sequential SL, where most clients remain idle while waiting for their turn, PSL enables clients to train in parallel. The server aggregates the smashed data uploaded by all clients to update the server-side model and simultaneously returns either personalized or globally aggregated gradients to each client. Clients then update their local models using the received gradients. This strategy effectively improves computational efficiency and reduces idle time among clients.

\noindent\textbf{Split federated learning (SFL)~\cite{joshi2021splitfed,gao2024pipesfl}.}
In SFL, all clients train their local parts of the model in parallel and update them based on their computed gradients. After a certain number of training rounds, each client sends its local model to a central federated server, which aggregates the models to form a global client model and then redistributes it. Unlike PSL, SFL introduces periodic global aggregation of client models, providing stronger global consistency but with additional synchronization overhead.
Specifically, SFL encompasses two server-side update variants: SFL-V1 updates the global trunk model by aggregating gradients from all clients via FedAvg, whereas SFL-V2 processes client activations sequentially in random order, executing immediate server model updates without global aggregation.

\noindent\textbf{Group-based split federated learning (GSFL)~\cite{liao2024parallelsfl,wu2023split,zhang2023cluster,Liu2025GSFLAP}.}
To address the efficiency bottlenecks and data heterogeneity inherent in standard SFL, GSFL restructures the flat client-server topology into a clustered architecture.
From a system perspective, clients with similar computational and network capabilities are grouped to mitigate the straggler effect and reduce synchronization delays. From a data perspective, clients with similar data distributions are clustered to counteract non-IID challenges, thereby alleviating client drift and preventing model divergence.

\noindent\textbf{Other framework variants.}
Based on the concept of split learning, several other novel frameworks have been proposed. RingSFL~\cite{shen2023ringsfl} achieves load balancing and effectively mitigates the bottleneck effect by constructing clients into a ring topology and adaptively distributing model layers based on each client's computational capabilities. 
HSFL~\cite{zhang2024splitllm,lin2025hierarchical,fan2025madrl} employs a hierarchical multi-server architecture (cloud–edge–device) to overcome the memory and communication bottlenecks of conventional single-layer split learning systems under large-scale user participation. PairingFL~\cite{yao2025pairingfl} lets paired clients train collaboratively, performing most updates within pairs to reduce server communication overhead. 
CLICOOPER~\cite{deng2026clicooper} introduces a serverless, multi-client SL framework for partially trusted environments, where a data-owning client delegates sequential model training to a chain of compute-contributing nodes.

\subsection{Split-based Pipeline Components}
\label{subsec:pipeline}

\begin{wrapfigure}{r}{0.63\textwidth}
    \centering
    \vskip -5pt
    \includegraphics[width=\linewidth]{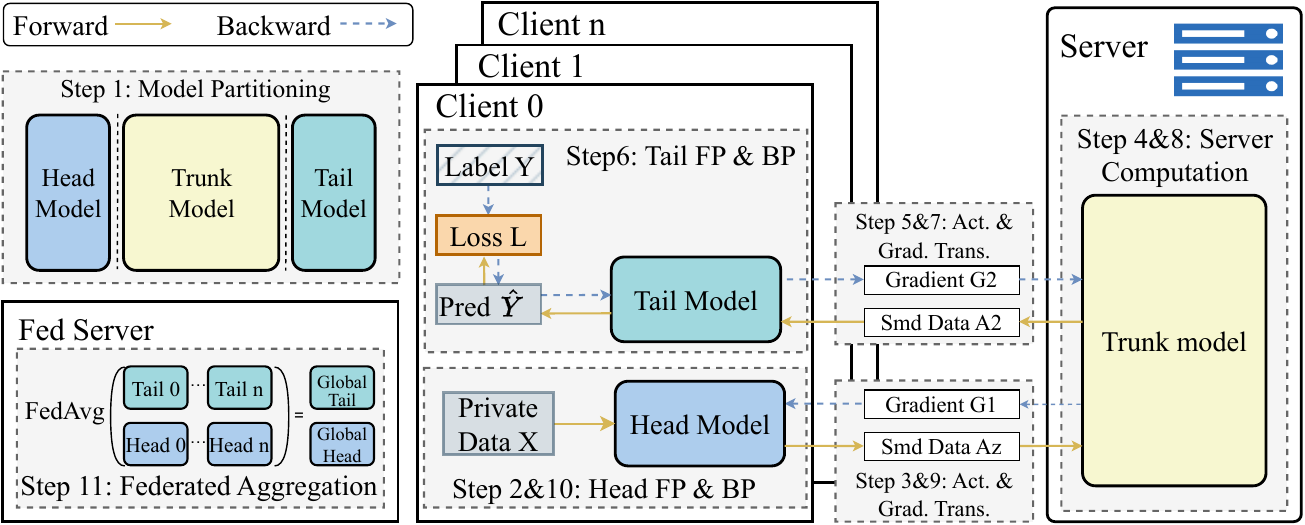}
    \vskip -5pt
    \caption{\lzh{Unified split-based LLM-FT pipeline.}}
    \vskip -5pt
    \label{fig:pipeline}
\end{wrapfigure}

In the split-based LLM-FT framework, model training relies on the coordinated operation of multiple critical components. Optimization efforts in this domain primarily focus on improving one or more of these components to enhance overall efficiency, performance, and privacy. This section first introduces a unified pipeline, as illustrated in Figure \ref{fig:pipeline}, which outlines the complete training workflow. Subsequently, we present a detailed analysis of the key components within this pipeline, providing a foundation for later discussions on component-level optimization strategies.
\lzh{In this unified pipeline, steps 1–4 and 9–10 constitute the core workflow shared by all split-based LLM-FT frameworks. To clearly delineate framework-specific operations, we append a dagger ($\dagger$) to steps 5–8 to denote procedures exclusive to the three-part split (u-shaped SL) architecture, and a hash symbol (\#) to step 11 to indicate the global synchronization mechanism uniquely required by split federated learning (SFL) paradigms.}

\noindent\textbf{Step 1: model partition.}
The complete LLM is typically partitioned at the granularity of embedding layers or individual transformer blocks. Depending on the architecture, it is divided into either two components (head model and trunk model for vanilla SL) or three components (head model, trunk model, and tail model for u-shaped SL). These components are then deployed across the client and server sides according to the adopted split strategy.

\noindent\textbf{Step 2: local head model forward propagation.}
The client performs forward propagation on the local head model using its local dataset, generating head model's output intermediate activations.

\noindent\textbf{Step 3: intermediate activation transmission (client to server).}
The client transmits the intermediate activations produced by the head model to the server over the network.

\noindent\textbf{Step 4: server-side processing.}
Upon receiving the intermediate activations, the server continues the forward propagation through the trunk model. In vanilla SL, the server directly computes the loss using transmitted labels and initiates backpropagation. Conversely, in u-shaped SL, the server suspends the process after the forward pass, awaiting gradients from the client. In both scenarios, the server ultimately computes the gradients corresponding to the intermediate activations output by the head model.

\noindent\textbf{Step 5$\dagger$: intermediate activation transmission (server to client).}
Because the server in a u-shaped SL architecture cannot directly access the original labels to compute the loss, it must transmit the intermediate activations produced by the trunk model back to the client for further processing.

\noindent\textbf{Step 6$\dagger$: local tail model forward and backward propagation.}
After receiving the intermediate activations from the server, the client performs forward propagation through its local tail model, computes the loss using local labels, and executes backpropagation to obtain the gradients with respect to the trunk model’s output activations.

\noindent\textbf{Step 7$\dagger$: intermediate gradient transmission (client to server).}
The client transmits the gradients of the trunk model’s output activations to the server over the network.

\noindent\textbf{Step 8$\dagger$: server-side trunk model backward propagation.}
Upon receiving the gradients from the client, the server continues the backward propagation through the trunk model to compute the gradients corresponding to the head model’s output activations.

\noindent\textbf{Step 9: intermediate gradient transmission (server to client).}
The server sends the gradients of the head model’s output activations back to the client.

\noindent\textbf{Step 10: local head model backward propagation.}
After receiving the gradients of the head model’s output activations, the client performs backward propagation on the local head model to update its parameters.

\noindent\textbf{Step 11\#: federated aggregation (for SFL-based frameworks).}
After several local training rounds, each client uploads its local model parameters to the federated server for global aggregation. The federated server aggregates these local models into a global model and redistributes it to all participating clients for the next training round.

\lzh{\subsection{Mapping Pipeline Components to Core Bottlenecks}}
\lzh{While the unified pipeline (steps 1--11) establishes a functional paradigm for split-based LLM-FT, mapping its sequential execution onto a distributed, heterogeneous, and potentially untrusted environment exposes critical vulnerabilities during actual deployment. By analyzing each step, we identified bottlenecks across three distinct dimensions, which also dictate the taxonomy of our subsequent sections:}

\begin{itemize}
    \item \textbf{Model-level bottlenecks.} In multi-client scenarios, executing local forward and backward passes (steps 2, 6, 10) on highly skewed, Non-IID datasets generates biased smashed data. When aggregated at the server or during global synchronization (step 11), this heterogeneity, coupled with server-client update imbalances, severely degrades model convergence and generalization. Furthermore, in sequential training or scenarios with partial client participation, the model is prone to catastrophic forgetting, inadvertently overwriting previously learned representations while optimizing for the distinct data distributions of current participants. Section 3 explores advanced strategies to rectify these performance degradations.
    
    \item \textbf{System-level bottlenecks.} At the initiation of the pipeline (step 1), the selection of the split point dictates the local memory footprint and profoundly impacts training latency and energy consumption due to client-server computational disparities. Subsequently, the reliance on continuous, bidirectional transmission of massive intermediate activations (steps 3, 5) and gradients (steps 7, 9) induces severe communication overhead. Furthermore, the central server (step 4) easily becomes a computational bottleneck when coordinating heterogeneous clients. Finally, frameworks that necessitate periodic global model aggregation (step 11) introduce rigid synchronization barriers. Under system heterogeneity, this leads to the straggler problem, where high-capacity clients experience significant training delays waiting for the slowest devices. Section 4 systematically reviews system-level optimizations designed to alleviate these structural inefficiencies.
    
    \item \textbf{Privacy and security vulnerabilities.} Although the pipeline inherently avoids the direct exchange of raw inputs and labels, the transmitted intermediate representations remain highly vulnerable. Specifically, a malicious server can exploit the intercepted smashed data (steps 3, 5) to launch smashed data inversion attacks, and it can utilize the backpropagated gradients (steps 7, 9) to execute gradient-based label reconstruction. Furthermore, because the server maintains control over a substantial segment of the model architecture and training process (step 4), it can actively manipulate the training process to initiate data poisoning and backdoor attacks. To counter these vulnerabilities, Section 5 details the prevailing threat landscape and systematically summarizes the corresponding privacy-preserving defense mechanisms.
    
\end{itemize}

\section{Model-Level Optimization and Performance Enhancement}
\label{section: model}
Model-level optimizations are pivotal for ensuring the effectiveness of Split-based LLM Fine-tuning, particularly in terms of convergence, accuracy, and robustness. Theoretically, in a standard single-client, single-server architecture, the complete training process of split learning is mathematically equivalent to centralized training, yielding nearly identical model performance. However, when scaled to collaborative fine-tuning scenarios involving multiple clients, this equivalence is fundamentally compromised. Specifically, three primary challenges emerge that undermine model performance:

\begin{itemize}
    \item \textbf{Challenge 1: data heterogeneity (non-IID).} The private data of each client often exhibits divergent distributions. The intermediate features generated from these non-IID datasets can bias the convergence trajectory of the server-side model, degrading global model accuracy.
    
    \item \textbf{Challenge 2: server-client update imbalance.} The server-side model processes aggregated smashed data from multiple clients, effectively learning from a ''larger batch'' than any single client. This causes the server to update faster than client-side models, leading to desynchronized convergence.
    
    \item \textbf{Challenge 3: catastrophic forgetting.} In sequential training or scenarios with partial client participation, the model tends to inadvertently overwrite representations learned from previous clients while optimizing for the distinct data distributions of current participants.
\end{itemize}

This section reviews recent methodologies designed to address these challenges through personalization, feature distribution alignment, and advanced optimization strategies. A comprehensive summary of these model-level optimization strategies is presented in Table~\ref{tab:model_optimization}. Although several foundational methods lack empirical validation on native LLM tasks, we have theoretically evaluated their algorithmic compatibility. Based on their core mechanisms, we project their suitable LLM applications and categorize them into Natural Language Understanding (NLU) and Generation (NLG), as detailed in the \emph{Applicabiliy} column.

\begin{table*}[t]
\centering
\vskip -1pt
\caption{Summary of model-level optimization and performance enhancement methods in split-based Learning. LLM Exp. indicates whether the original paper empirically evaluated the method on LLM datasets ($\checkmark$). Applicability represents our theoretical assessment of the method's viable scope in the LLM era, categorized into Natural Language Understanding (NLU) and Generation (NLG) tasks.}
\vskip -10pt
\label{tab:model_optimization}
\resizebox{\textwidth}{!}{%
\setlength{\tabcolsep}{3pt}
\begin{tabular}{c|c c c c c p{9.5cm}}
\toprule
\textbf{Target Challenge} & \textbf{Approach} & \textbf{Method} & \textbf{Year} & \textbf{LLM Exp.} & \textbf{Applicability} & \multicolumn{1}{c}{\textbf{Key Innovation / Contribution}} \\
\midrule

\multirow{15}{*}{\makecell[c]{\textbf{Challenge 1:} \\ \textbf{Data Heterogeneity}}} 
 & \multirow{7}{*}{\makecell[c]{\textbf{Personalization} \\ \textbf{\& Generalization}}} 
 & SplitGP~\cite{han2023splitgp} & 2023 & - & NLU+G & Dual-task loss for personalization/generalization; $\lambda$-based aggregation. \\
 & & PFSL~\cite{wadhwa2023pfsl} & 2023 & - & NLU+G & Two-stage training with Work Fairness Constraint for load balancing. \\
 & & DualFed~\cite{zhu2024dualfed} & 2024 & - & NLU+G & Dual-feature representation with separate encoders for global/local tasks. \\
 & & FSMKD~\cite{luo2024fsmkd} & 2024 & - & NLU+G & Two-body structure with Mutual Knowledge Distillation (soft labels). \\
 & & ESL~\cite{chawla2024beyond} & 2024 & - & NLU+G & Two-stage training targeting non-IID data optimization. \\
 & & SplitLPF~\cite{chen2023personalized} & 2024 & - & NLU+G & Contribution-based fair aggregation for U-shaped structures. \\
 & & \llm{FlexP-SFL}~\cite{yuan2025flexpsfl} & 2025 & \checkmark & NLU+G & KL divergence regularization to align intermediate representations. \\
\cmidrule(lr){2-7} 
 & \multirow{6}{*}{\makecell[c]{\textbf{Smashed Data} \\ \textbf{Aggregation and} \\ \textbf{Scheduling}}}
 & MiniBatch-SFL~\cite{huang2023minibatch} & 2023 & - & NLU+G & Centralized gradient averaging of smashed data to remove client drift. \\
 & & MergeSFL~\cite{liao2024mergesfl} & 2024 & - & NLU & Feature merging for non-IID; Joint optimization with genetic algorithms. \\
 & & SCALA~\cite{yang2024scala} & 2024 & - & NLU & Concatenated activations with Logit Adjustment for long-tail data. \\
 & & S2FL~\cite{yan2023s2fl} & 2024 & - & NLU & Re-grouping server-side features by label to simulate IID batches. \\
 & & Hourglass~\cite{he2025hourglass} & 2025 & - & NLU+G & LSH-based ''dissimilar priority'' scheduling for asynchronous processing. \\
 & & Ampere~\cite{zhang2025ampere} & 2025 & - & NLU+G & Unified activation merging to reduce heterogeneity. \\
 \cmidrule(lr){2-7} 
 & \multirow{8}{*}{\makecell[c]{\textbf{Client Clustering} \\ \textbf{and Pairing}}}
 & CSFL~\cite{xu2024csfl} & 2024 & - & NLU & Clustering clients based on label distribution similarity and adaptively applying global or personalized aggregation. \\
 & & \llm{ParallelSFL}~\cite{liao2024parallelsfl} & 2024 & \checkmark & NLU & Clustering clients to ensure the assembled data distribution within groups closely approximates the global IID. \\
 & & PairingFL~\cite{yao2025pairingfl} & 2025 & - & NLU & Selecting clients with dissimilar and diverse data distributions to maximize marginal gain for each training round. \\
 & & GSFL~\cite{Liu2025GSFLAP} & 2025 & - & NLU+G & Grouping devices with similar data distributions measured via cosine similarity on encrypted intermediate features. \\
\midrule

\multirow{5}{*}{\makecell[c]{\textbf{Challenge 2:} \\ \textbf{Update Imbalance}}}
 & \multirow{2}{*}{\makecell[c]{\textbf{Learning Rate} \\ \textbf{Scaling}}}
 & SGLR~\cite{pal2021server} & 2022 & - & NLU+G & Server learning rate acceleration based on client count power scaling. \\
 & & musfl~\cite{liang2025towards} & 2022 & - & NLU+G & Coordinated Local-Global Learning Rate Compensation. \\
\cmidrule(lr){2-7} 
 & \multirow{4}{*}{\makecell[c]{\textbf{Smashed Data} \\ \textbf{Mixup}}}
 & LocFedMix-SL~\cite{oh2022locfedmix} & 2022 & - & NLU & Smashed data mixing and mutual information regularization. \\
 & & Mix2SFL~\cite{oh2023mix2sfl} & 2024 & - & NLU & Two-way manifold mixup on paired smashed data. \\
 & & GAS~\cite{yang2025gas} & 2025 & - & NLU & Generative Gaussian distribution for virtual activation sampling. \\
 & & SMixSL~\cite{tinh2025smixsl} & 2025 & - & NLU & Cutout, CutMix, and Mixup applied to smashed features. \\
\midrule

\multirow{2}{*}{\makecell[c]{\textbf{Challenge 3:} \\ \textbf{Catastrophic Forgetting}}}
 & \multirow{2}{*}{\makecell[c]{\textbf{Knowledge} \\ \textbf{Retention}}} 
 & MultiSFL~\cite{xia2025multisfl} & 2025 & - & NLU+G & Helper clients extract and replay missing knowledge. \\
 & & SLwF~\cite{feng2024slwf} & 2025 & - & NLU+G & Review/Distinguish/Decelerate modules for continual learning. \\
\bottomrule
\end{tabular}
}
\vskip -11pt
\end{table*}

\subsection{Mitigating Data Heterogeneity (Non-IID)}
The inherent statistical divergence among client datasets poses a severe threat to the convergence and generalization of the global model. To address the feature shifts induced by non-independent and identically distributed (Non-IID) data, recent literature predominantly explores three main strategies: decoupling personalization and generalization via architectural design, smashed data aggregation and scheduling, and client clustering and pairing.

\subsubsection{\lzh{Decoupling personalization and generalization via architectural design}}

In scenarios where client data exhibits non-IID characteristics, client-side models are updated exclusively on skewed local datasets. Consequently, the generated intermediate features often lack representativeness for global classes (or fail to capture meaningful information for out-of-distribution categories). This feature shift not only impedes the effective generalization of the server-side model but also renders a simple globally aggregated model insufficient for adapting to the personalized distributions of individual clients. Consequently, recent methods~\cite{han2023splitgp, chen2023personalized, zhu2024dualfed, luo2024fsmkd, chawla2024beyond, wadhwa2023pfsl,yuan2025flexpsfl} focus on reconciling the conflict between personalization and generalization in SFL scenarios. SplitGP~\cite{han2023splitgp} decouples the model into a client-side component dedicated to strong personalization and a server-side component focused on strong generalization. By jointly optimizing both components via a multi-exit objective function and a hybrid aggregation strategy, the framework achieves robustness to both local and out-of-distribution data while significantly reducing inference latency. Similarly, DualFed~\cite{zhu2024dualfed} addresses this dilemma by exploiting the hierarchical nature of deep representations to decouple generalization from personalization. Instead of structurally splitting the model across devices, it inserts a personalized projection network between the shared encoder and the classifier. This architecture yields dual representations: pre-projection features, which retain generalized information to support a global classifier, and post-projection features, which are mapped to a personalized space for local adaptation. To prevent the optimization of local tasks from compromising the generalized features, it employs a stage-wise training strategy that alternates between updating the personalized main branch and the global classifier.
SFML~\cite{xia2025sfml} maintains a dedicated privacy model and the split segments (head and tail) of a public model locally on each client, and leverages a mutual learning module to enable bidirectional knowledge transfer between the two models, perfectly balancing lightweight computation, personalization, and global generalization.

\subsubsection{Smashed data aggregation and scheduling}

\noindent Beyond balancing personalization and generalization, another effective paradigm for mitigating client-side data heterogeneity is the direct regulation of intermediate activations. By merging or scheduling the smashed data uploaded by different clients, the server can smooth out the statistical variance between batches. This ensures that the server-side model observes a more representative and balanced global distribution during training, preventing it from overfitting to any single client's skewed domain.

\noindent\textbf{Smashed data aggregation.}
Some methods~\cite{huang2023minibatch, liao2024mergesfl, yang2024scala, yan2023s2fl, zhang2025ampere} use Feature Merging to make the server-side feature batches approximate the IID distribution of feature batches.
MergeSFL~\cite{liao2024mergesfl} addresses statistical heterogeneity by constructing a mixed feature sequence that approximates an IID distribution. specifically, the server dynamically selects workers under bandwidth constraints to minimize the KL divergence of the combined data. By merging features from these selected workers, the framework rectifies the top model's update direction, significantly enhancing final accuracy. SCALA~\cite{yang2024scala} simulates centralized training to mitigate local label skew by concatenating client-side activations at the server. Furthermore, it addresses the global long-tail distribution caused by partial participation through logit-adjusted loss functions, thereby effectively improving SFL performance on Non-IID data.
\lzh{However, applying these strategies to native LLMs introduces severe memory bottlenecks. Given the massive size of LLM activations, concatenating smashed data from multiple clients rapidly exhausts server GPU capacity, triggering catastrophic out-of-memory failures.}

\noindent\textbf{Smashed data scheduling.} 
Taking a data-content-centric approach, Hourglass~\cite{he2025hourglass} introduces a ''Dissimilar Priority Scheduling'' (DPS) mechanism. It utilizes Locality-Sensitive Hashing (LSH) to efficiently identify intermediate features that differ significantly from those currently being processed. By prioritizing these ''dissimilar'' features during server-side training, Hourglass maximizes the information gain of each update step, thereby accelerating convergence even under severe data heterogeneity.

\subsubsection{Client clustering and pairing} 

\noindent Smashed data aggregation and scheduling alleviate data heterogeneity after the server receives the intermediate features, while client clustering and pairing preemptively reorganize the collaborative structure based on statistical characteristics, thereby making the training data within a group approximate an IID distribution.
To eliminate the impact of non-IID data on model accuracy, ParallelSFL~\cite{liao2024parallelsfl} partitions participating workers into multiple clusters , requiring the assembled data distribution of each cluster to closely approximate the global IID distribution. It introduces KL-divergence to measure the distribution gap. The algorithm first uses K-means to group similar workers , then greedily selects workers across these groups to build clusters , and finally fine-tunes by exchanging workers across clusters to minimize the KL-divergence of each cluster. PairingFL~\cite{yao2025pairingfl} adopts a finer-grained collaboration strategy. Specifically, while ensuring participation fairness across all clients , it employs a greedy algorithm to iteratively select clients with the maximum marginal gain (i.e., those with the most dissimilar and diverse data distributions) to form the training subset for each round. GSFL~\cite{Liu2025GSFLAP} uses a lightweight hash encryption function to encrypt the intermediate features of devices and sends them to the server. The server uses Cosine Similarity to measure the similarity of data distributions among different devices, and assigns devices with similar data distributions into the same group, thereby mitigating the negative impact brought by Non-IID data. Similarly, CSFL~\cite{xu2024csfl} evaluates the similarity of clients' label distributions using cosine similarity, and subsequently employs the K-means algorithm to partition them into $K$ clusters. During the training phase, the framework adaptively toggles between a global aggregation strategy and a personalized cluster-specific approach depending on the degree of inter-cluster heterogeneity.

\lzh{\subsection{Bridging Server-Client Update Imbalance}}
\lzh{\noindent In multi-client scenarios, the server effectively processes a dynamically expanding batch aggregated from concurrent participants, leading to a profound desynchronization between server and client learning trajectories. To harmonize this update imbalance, optimization efforts have focused on feature-level augmentation and learning rate scaling.}
\subsubsection{Smashed data mixup}

\noindent In multi-client parallel split learning architectures (e.g., SFL, PSL), the server-side model processes aggregated smashed data from $n$ concurrent clients in each communication round. Consequently, the server-side model experiences an effective learning rate that is amplified by a factor of $n$ relative to the client-side models. This server-client Update Imbalance induces training inconsistencies, as a higher effective learning rate typically necessitates a proportionally larger batch size to preserve the stability of gradient estimation. To mitigate this discrepancy, several approaches~\cite{oh2022locfedmix, oh2023mix2sfl, yang2025gas, tinh2025smixsl} employ data augmentation directly on smashed data to harmonize the learning speeds of the clients and the server. LocFedMix-SL~\cite{oh2022locfedmix} proposes applying the Mixup technique at the level of intermediate activations. By synthesizing augmented smashed data via linear interpolation, this method effectively expands the server's virtual batch size to align with its amplified learning rate, thereby stabilizing the system's training dynamics. Building on this, Oh et al.~\cite{oh2023mix2sfl} proposed Mix2SFL, a framework that synchronizes forward and backward flows via a "two-way mixup" strategy. Specifically, it employs SmashMix to combinatorially mix smashed data, balancing the server-client update ratio with averaged forward flows. Furthermore, Mix2SFL incorporates GradMix to average split-layer gradients, which not only aligns backward signals but also enhances communication efficiency by enabling downlink broadcasting.
\lzh{While potentially viable for natural language understanding (NLU) tasks, this strategy is fundamentally incompatible with natural language generation (NLG) tasks, as interpolating variable-length hidden states and shifted next-token labels destroys semantic integrity and triggers convergence collapse.}

\subsubsection{Learning rate scaling}

\noindent Due to the server-side large effective batch problem, where the effective batch size at the server is the sum of all client batches resulting in an update imbalance between the server and the clients, SGLR~\cite{pal2021server} splits the learning rate into server-side and client-side components and independently accelerates the server-side learning rate. MU-SplitFed~\cite{liang2025towards} binds the server learning rate to the unbalanced step count $\tau$ and reduces it to suppress the variance and client drift caused by stale information , while simultaneously amplifying the global aggregation learning rate to compensate for the convergence slowdown caused by the shortened local steps.

\subsection{Overcoming Catastrophic Forgetting}
\noindent In sequential training or scenarios with partial client participation, the model is highly susceptible to overwriting previously acquired representations when adapting to newly introduced skewed distributions.

\subsubsection{Knowledge retention}

\noindent In vanilla SL, the model is trained sequentially across multiple clients with Non-IID local data, which often causes it to adapt to the current device while losing consistency with previously learned knowledge, resulting in degraded performance.
To mitigate this issue, SLwF~\cite{feng2024slwf} incorporates three lightweight modules: a review module that revisits prior knowledge through contrastive learning, a dividing module that facilitates learning of new or hard classes, and a decelerated update module that stabilizes model updates via an EMA-based strategy. Together, these modules enable SLwF to better balance knowledge retention and adaptation during client switching.
In the SFL scenario, if only a portion of the clients participate in each training round, the system may also be prone to catastrophic forgetting and will also be challenged by gradient divergence caused by non-independent and non-IID data. To address these issues, MultiSFL~\cite{xia2025multisfl} introduces a multi-model aggregation architecture that maintains multiple branch models to promote knowledge sharing. It further mitigates forgetting through a score-vector-based knowledge replay mechanism that leverages idle clients to compensate for missing-category features. Moreover, MultiSFL employs the Federated Gradient Norm (FGN) to identify critical learning periods and adaptively adjust the sampling proportion, thereby balancing model performance and communication overhead. 
\section{System-Level Optimizations}
\label{section:system}
\begin{wrapfigure}{r}{0.52\textwidth}
    \centering
    \vskip -15pt
    \includegraphics[width=\linewidth]{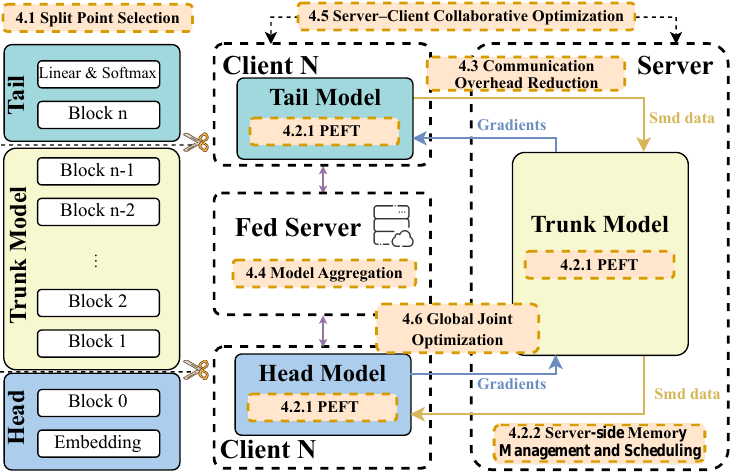}
    \caption{Overview of system-level optimization for Split-based LLM-FT}
    \vskip -5pt
    \label{fig:Overview of System-Level Optimization}
\end{wrapfigure}

System-level optimizations are critical for improving the overall efficiency, scalability, and stability of Split-based LLM Fine-Tuning (LLM-FT) frameworks. 
While model partitioning effectively reduces local computational burdens, it introduces new system challenges such as high communication overhead, unbalanced resource utilization, and synchronization delays. 
This section provides a comprehensive review of recent efforts aimed at optimizing the system architecture and execution efficiency of Split-based LLM-FT. 
To provide a holistic view, Fig.~\ref{fig:Overview of System-Level Optimization} illustrates the integration of these strategies within the unified training pipeline \lzh{introduced in Subsection~\ref{subsec:pipeline}.} 
\lzh{
Guided by the execution flow of this pipeline, the following subsections sequentially detail split point selection, computation optimization and resource management, communication overhead reduction, and model aggregation strategies. Finally, the section concludes by exploring server–client collaborative optimization and global joint optimization in the last two subsections.
Notably, throughout this section, works denoted with an asterisk (*) are either specifically designed for or experimentally validated on LLMs. Conversely, unmarked methods originate from conventional split learning tasks, yet their underlying principles exhibit strong transferability to LLM fine-tuning scenarios.
}

\subsection{Split Point Selection - Step 1}
The split point determines how many layers are computed locally, which must fit within the client’s maximum GPU memory. Due to differences in computing power between client and server, the chosen split also affects training latency and energy consumption. 
Recent studies have explored strategies to optimize split-point selection—for example, by minimizing the overall training delay or reducing energy consumption during distributed training~\cite{samikwa2022ares,marinova2025optimal,trung2025latency}.
Additionally, in scenarios involving multiple concurrent clients (such as SFL), each client can select different split points based on its own computational capacity~\cite{chen2025memory}. This approach adapts to individual computational capabilities while balancing training times across clients, thereby mitigating the low training efficiency caused by stragglers~\cite{yan2023s2fl,ma2025splitfrozen,wu2025sfl}. 
\lzh{Importantly, a critical distinction exists in native LLMs: unlike CNNs, Transformer blocks maintain a strictly constant hidden dimension. Consequently, shifting the split point does not alter the communication payload size. Therefore, dedicated communication reduction techniques are required to minimize this overhead.}

\subsection{Computation scheduling and memory managing - Step 2 \& 4 \& 6$\dagger$ \& 8$\dagger$}
While model partitioning fundamentally mitigates the burden on individual clients, the sheer scale of LLMs continues to impose severe computational and storage overheads across the entire collaborative system. Consequently, optimizing computation scheduling and memory management requires a comprehensive, two-pronged approach. First, to further minimize the overall memory footprint and computational costs, recent research extensively integrates parameter-efficient architectures (PEFT) across both client and server models. Second, as the pivotal hub executing the Trunk Model and managing concurrent multi-client requests, the central server requires dedicated resource orchestration. Without efficient scheduling and memory management, the server is highly susceptible to critical system bottlenecks, primarily driven by concurrent GPU memory contention and the {\em straggler effect}.

\subsubsection{Parameter-efficient fine-tuning (PEFT)}
The direct application of traditional full fine-tuning (FFT) in split learning presents significant challenges due to severe memory and communication bottlenecks. On the one hand, FFT necessitates maintaining full-parameter optimizer states and gradients, imposing prohibitive memory overheads on both resource-constrained clients and the central server. On the other hand, in collaborative training scenarios, the requirement to aggregate the complete model parameters, rather than lightweight updates, significantly exacerbates communication costs. To mitigate these limitations, recent research has extensively integrated parameter-efficient fine-tuning (PEFT) techniques, including low-rank adaptation (LoRA)~\cite{lin2024splitlora, lin2025hsplitlora, zhao2025efficient, ma2025splitfrozen, huang2025sldp}, adapters~\cite{borzunov2023petals, hu2024menos}, and prompt tuning~\cite{borzunov2023petals, cao2024sfprompt,gu2026del}, into the split learning paradigm.
SplitLoRA$^*$~\cite{lin2024splitlora} is the first SFL framework designed for LLM fine-tuning. By integrating LoRA with aplit federated learning (SFL), it aims to address the computation and communication bottlenecks faced by traditional federated learning on resource-constrained edge devices, utilizing periodic client adapter aggregation to maintain the benefits of parallel training.
To accommodate heterogeneous computing environments, HSplitLoRA$^*$~\cite{lin2025hsplitlora} proposes a dynamic framework that adapts model split points and LoRA ranks to individual client budgets. It further enhances efficiency through an important weight identification scheme and ensures robust model updates via a noise-free adapter aggregation mechanism using matrix concatenation.
Furthermore, Menos$^*$~\cite{hu2024menos} targets the scalability bottleneck on the server side. To mitigate server GPU memory overloads, Menos exploits the frozen nature of base model parameters in adapter-based fine-tuning. It implements a spatial sharing mechanism where a single copy of the base model is shared across multiple clients, thereby eliminating the redundant memory consumption of duplicating massive model parameters for concurrent tasks.

\subsubsection{Server-side memory management and computation scheduling}
\lzh{Unlike clients, which handle only local forward and backward propagation, the central server must process massive intermediate data and coordinate multiple concurrent clients. Consequently, the server frequently emerges as the system's computational and memory bottleneck. To mitigate this, recent studies focus on server-side optimizations, specifically memory sharing and priority-based scheduling.}

\noindent\textbf{Server memory management and sharing.}
Fine-tuning LLMs places a tremendous burden on server GPU memory, especially when maintaining unique states (e.g., optimizer states, gradients) for multiple concurrent clients.
\llm{Menos$^*$~\cite{hu2024menos}} mitigates this via a dynamic memory sharing mechanism. By swapping client-specific contexts (e.g., LoRA adapters) to CPU memory during idle periods while sharing a single frozen base LLM on the GPU, it significantly reduces memory footprint and prevents OOM errors under high concurrency. Similarly, Hourglass~\cite{he2025hourglass} optimizes resource utilization by decoupling server-side model partitions from the client count. Instead of assigning a unique partition per client, it employs $M$ partitions matching the available GPUs to enable efficient data parallelism. This approach processes client features in groups, eliminating the overhead of frequent partition swapping and accelerating model convergence.

\noindent\textbf{Server priority-based scheduling.}
In split-based learning frameworks, the server acts as a centralized hub, continuously receiving forward and backward propagation tasks from multiple clients. Due to the inherent heterogeneity in clients' computing power and network bandwidth, the arrival times of these tasks are highly irregular. Consequently, a standard First-In-First-Out (FIFO) scheduling strategy often leads to suboptimal execution orders, causing significant server idle time and pipeline bubbles. To address this, priority-based scheduling aims to reorder the execution of tasks to maximize pipeline parallelism. \llm{PipeSFL$^*$~\cite{gao2024pipesfl}} introduces a mechanism based on clients' overall lag, scheduling stragglers earlier to hide communication latency. Similarly, \llm{Chen et al.$^*$~\cite{chen2025memory}} employ a greedy algorithm that prioritizes tasks with longer backward propagation times to optimize server-side sequential training. Beyond heuristics, Tirana et al.~\cite{tirana2024workflow} formulate workflow orchestration as a joint client-helper assignment and scheduling problem, proving it is NP-hard and solving it via an ADMM-based decomposition to minimize the training makespan.

\subsection{Communication Overhead Reduction - Step 3 \& 5$\dagger$ \& 7$\dagger$ \& 9}
Due to the frequent transmission of activation values and gradients between the client and server, communication overhead accounts for a significant portion of the time spent in SL training. 
\lzh{In the context of LLMs, this bottleneck is severely exacerbated by the massive dimensionality of their hidden states.} 
To mitigate this excessive communication overhead, recent optimization efforts generally fall into three strategies: (1) activation and gradient compression, (2) communication frequency reduction, and (3) gradient aggregation.

\subsubsection{Activation and gradient compression}
In Split-based LLM-FT, activation and gradient compression has emerged as the most direct adopted approach to reduce communication overhead. 
By compressing the intermediate activations transmitted from the client to the server and the gradients sent back during backpropagation, these methods effectively decrease the data volume per iteration and shorten transmission latency. 
However, compression inevitably leads to partial information loss, which can adversely affect model convergence and overall performance if not properly compensated. 
Therefore, designing efficient compression techniques that strike a balance between communication efficiency and model accuracy has become a major research focus.
Existing activation and gradient compression methods can be broadly categorized into three primary streams: \textbf{top-$k$ compression}, \textbf{dimensionality reduction}, and \textbf{quantization}, alongside \textbf{other emerging techniques} (e.g., SVD and Token-based methods). While activation and gradient compression methods were initially pioneered for traditional SL frameworks, they provide foundational theoretical blueprints for mitigating the massive communication bottlenecks in Split-based LLM-FT.

\noindent\textbf{Top-$k$ compression} transmits only the most significant $k$ elements of the activation or gradient tensors, thereby reducing the transmitted data while preserving key information. RandTopk~\cite{zheng2023reducing} introduces randomness into Top-k selection, enabling the model to better explore the feature space, avoid local minima, and achieve higher performance and stability under the same compression ratio. MS~\cite{zhou2024mask} compensates for lost values in Top-k sparsification with mask encoding, reducing compression errors and gradient bias to improve Split Learning convergence. FedDFC~\cite{gao2025communication} dynamically adjusts the Top-k compression ratio for each client based on their computing capacity and network conditions, optimizing communication efficiency and reducing waiting time.

\noindent\textbf{Dimensionality reduction}~\cite{ayad2021improving,shao2020bottlenet++,shu2024dynsplit} utilizes a client-side encoder to map intermediate features into low-dimensional representations and a server-side decoder for reconstruction, significantly lowering communication overhead. BottleNet++~\cite{shao2020bottlenet++} incorporates joint source–channel coding, achieving high compression rates by exploiting DNN tolerance to feature distortion. DynSplit~\cite{shu2024dynsplit} employs a multi-path design where a gating network dynamically selects compression routes based on intermediate features, balancing accuracy and communication efficiency. 
Building upon dimensionality reduction, \llm{DEL$^*$~\cite{gu2026del}} incorporates stochastic n-bit quantization to further reduce communication overhead and ensure strict differential privacy , while leveraging server-side soft prompts to compensate for the performance degradation induced by privacy noise.

\noindent\textbf{Quantization}~\cite{chen2021communication,wang2022fedlite,oh2025communication,lin2025slacc,pham2023binarizing} minimizes transmission costs by representing intermediate data with reduced bit-width precision (e.g., 8-bit integers) instead of standard floating-point formats.  FedLite~\cite{wang2022fedlite} further reduces communication by compressing activations via product quantization and mitigating the resulting noise with a gradient correction scheme, achieving high compression ratios with minimal accuracy loss. SplitFC~\cite{oh2025communication} leverages feature dispersion for adaptive compression, integrating standard deviation-based feature-wise dropout with a quantization scheme sensitive to feature variability. SL-ACC~\cite{lin2025slacc} employs Adaptive Channel Importance Identification (ACII) to quantify channel contributions via Shannon entropy and utilizes Channel Grouping Compression (CGC) to cluster channels for adaptive quantization bit-width allocation.
By applying extreme quantization (binarization) to local client layers, combined with leakage restriction and differential privacy , B-SL~\cite{pham2023binarizing} drastically reduces edge computation and communication overheads while maintaining high accuracy. However, the efficacy of such extreme quantization in LLM fine-tuning remains an open question.

\noindent\textbf{Other techniques.} Beyond the aforementioned categories, several distinct approaches exist. C3-SL~\cite{hsieh2022c3} employs circular convolution for batch-wise compression, superimposing multiple quasi-orthogonal samples into a single vector to reduce communication overhead while retaining feature details. SVD-SFT~\cite{shi2022efficient} applies Singular Value Decomposition (SVD) to decompose the split-layer feed-forward network (FFN) into low-rank matrices, transmitting only compressed activations to significantly reduce communication overhead. \llm{StreamSL$^*$~\cite{radovivc2024train}} focuses on transmitting only unpadded token embeddings and reconstructing the complete sequence using zero padding at the receiving end, thereby achieving lossless embedding compression and improving sequence-level efficiency.

\subsubsection{Communication frequency reduction}
Complementary to compression techniques that minimize the data volume per iteration, another research stream focuses on reducing the frequency of client-server synchronization. By decreasing the total number of communication rounds, these methods effectively lower the aggregate communication overhead and latency. 
Specifically, these approaches primarily encompass three strategies: \textbf{intermittent local model update}~\cite{ayad2021improving,chen2021communication,mu2023CSEFSL}, \textbf{local loss updates}~\cite{han2021accelerating,mu2023CSEFSL,nairfsl,zhang2025ampere, chopra2021adasplit,cao2024sfprompt,wu2025sfl,lin2025eslleo}, and \textbf{data pruning}~\cite{cao2024sfprompt,han2023splitgp}.

\noindent\textbf{Intermittent local model update.} 
This strategy relaxes the strict synchronization between the client and server, eliminating the necessity to exchange activations and gradients at every training iteration. Instead of continuous communication, these methods allow the system to skip transmission steps based on specific criteria.
MESL~\cite{ayad2021improving} employs an adaptive threshold mechanism during backward propagation, where the server withholds gradient transmission to the client if the current loss falls below a dynamic threshold. 
In contrast, CSE-FSL~\cite{mu2023CSEFSL} focuses on the forward pass by aggregating local updates for $h$ batches, thereby significantly reducing the frequency of smashed data transmission.

\noindent\textbf{Local loss updates.}
LocSplitFed~\cite{han2021accelerating} pioneered the acceleration of federated split learning through locally generated losses. By employing a lightweight auxiliary network, it enables clients to update their local models using locally generated losses without waiting for back-propagated gradients from the server, thereby achieving parallel training. This eliminates the transmission of back-gradients, drastically cutting down latency and bandwidth consumption. However, the lack of direct server supervision in prior auxiliary models often leads to inaccurate local gradients and performance degradation. Recent studies, such as FSL-SAGE~\cite{nairfsl} and ESL-LEO~\cite{lin2025eslleo}, address this by introducing a periodic alignment strategy where auxiliary models are regularly updated to accurately mimic the server's behavior. This synchronization bridges the gap between local training and global objectives, ensuring a theoretical convergence rate of $\mathcal{O}(1/\sqrt{T})$ while preserving communication efficiency.

\noindent\textbf{Data pruning.}
This strategy entails filtering out redundant data samples on the client side based on specific criteria before they enter the communication-intensive split training or inference pipeline.
\llm{SFPrompt$^*$~\cite{cao2024sfprompt} }implements a metric-based pruning mechanism using the Error L2 Norm (EL2N) score to quantify sample importance. By utilizing local model components to calculate these scores, the framework selectively retains only high-contribution samples for server interaction. This approach effectively filters out uninformative data, significantly reducing the volume of smashed data transmission while maintaining model performance.
Extending this to inference, SplitGP~\cite{han2023splitgp} minimizes offloading via confidence-based pruning. 
It utilizes Shannon entropy to quantify uncertainty, processing confident predictions locally and only transmitting high-entropy samples to the server. This dynamic filtering reserves bandwidth for difficult tasks necessitating server-side generalization.

\subsubsection{Gradient aggregation.}
Traditional SFL and PSL frameworks typically rely on independent point-to-point communication links between the server and each client to transmit gradients. This paradigm causes the downlink communication overhead to scale linearly with the number of clients, often resulting in severe communication bottlenecks. 
To address this limitation, several works~\cite{liang2025communication,oh2023mix2sfl,pal2021server} employ gradient aggregation strategies by consolidating gradients from multiple clients at the server. By broadcasting the unified result to all participants simultaneously, these approaches replace resource-intensive individual transmissions with a single efficient broadcast operation, effectively decoupling communication costs from the client population.
Specifically, EPSL~\cite{lin2024efficient} introduces a novel last-layer gradient aggregation mechanism, wherein the server aggregates gradients computed at the last layer of the server-side model before backpropagation. By incorporating an aggregation ratio and broadcasting the aggregated gradients to all participating clients, the framework effectively minimizes downlink communication latency, as it completely eliminates the linear scaling of transmission payload. 

\subsection{Model Aggregation Strategies - Step 11\#}
Compared to PSL, existing SFL frameworks necessitate the explicit synchronization and aggregation of client-side model weights after multiple local iterations. Beyond the computational and communication overheads, this synchronization imposes a rigid barrier. Under system heterogeneity, this leads to the straggler problem, causing significant training delays as high-capacity clients must wait for the slowest devices to synchronize. MHSL~\cite{joshi2021splitfed} modifies the SplitFedV2 architecture by eliminating the aggregation and synchronization of client-side models ($W_c$), effectively reverting to a PSL-like paradigm. While the resulting performance degradation is negligible for simple tasks and homogeneous settings, this approach suffers from significant accuracy loss when applied to complex datasets or heterogeneous scenarios.
To mitigate the resource inefficiency of synchronous client model aggregation, where fast devices idle waiting for stragglers, DT-ASFL~\cite{stephanie2023digital} employs an asynchronous aggregation strategy. This enables immediate processing of partial updates, allowing fast devices to train continuously. Furthermore, a "step-aware" mechanism addresses model staleness by weighting updates based on a delay variable; models derived from newer global parameters are assigned higher influence compared to those based on older versions. Similarly, SFL-LEO~\cite{wu2025sfl} proposes a staleness-aware asynchronous strategy to handle intermittent connectivity. It mitigates staleness through a dual mechanism: a soft penalty factor that progressively reduces the aggregation weight of lagged updates based on their delay extent, and a hard cutoff threshold that completely discards excessively stale gradients to prevent model divergence. In contrast, SAFSL~\cite{ao2025semi} adopts a semi-asynchronous protocol governed by dynamic time windows. By aggregating only updates completed within the window and allowing stragglers to continue training uninterrupted for future rounds, it effectively eliminates synchronization delays. Furthermore, it employs participation-interval-based weighting to balance contributions and prevent bias toward faster, more frequent devices.

\subsection{Server–Client Collaborative Optimization}
Traditional SL paradigms impose a strict serial dependency between the client and the server, where the client's forward propagation, data transmission, and the server's forward propagation must be executed sequentially. This "stop-and-wait" interaction mechanism causes computational nodes to remain idle for prolonged periods during communication, severely constraining system resource utilization and training throughput. 
To circumvent this bottleneck, Server-Client Collaborative Optimization aims to reconstruct the interaction protocol between the edge and the cloud. By employing strategies such as asynchronous training~\cite{chen2021communication, radovivc2024train, yang2025gas} and parallel training~\cite{zhang2024pipar, gao2024pipesfl, he2025hourglass, ma2025splitfrozen}, these approaches effectively achieve the decoupling and overlapping of computation and communication.

\subsubsection{Asynchronous training strategies.}
Departing from strict synchronization, AsyncSL~\cite{chen2021communication} implements a loss-based asynchronous update mechanism that dynamically adjusts the synchronization frequency between the client and server. By evaluating the training loss drop against a preset threshold, the framework determines whether to execute a full update or shift to an asynchronous 'silent' mode. This allows the server to continue training using stale activations while the client skips gradient reception and local updates, thereby achieving substantial reductions in both communication overhead and overall training latency.
StreamSL$^*$~\cite{radovivc2024train} achieves full computation-communication overlap by freezing client-side parameters, restricting local operations to the forward pass. This allows the client to asynchronously transmit intermediate features and labels, immediately processing subsequent batches without awaiting gradient returns.
GAS~\cite{yang2025gas} resolves the straggler issue in SFL by decoupling server and client operations via a dual-buffer mechanism. This architecture enables clients to upload data at their own pace while the server executes continuous, non-blocking updates. To mitigate the resulting bias where faster clients dominate the training, the server dynamically synthesizes virtual activations from historical distributions to balance the input stream.

\subsubsection{Parallel training strategies.}
To address the resource under-utilization caused by the sequential execution of computation and communication, PiPar~\cite{zhang2024pipar} introduces a pipeline parallelism framework that subdivides batches into mini-batches. This mechanism enables the concurrent overlapping of device-side and server-side propagation with the transmission of activations and gradients across mini-batches, thereby effectively minimizing idle time and significantly accelerating the training process.  To further enhance training efficiency under resource constraints, $C^2P^2SL$~\cite{liu2025communication} formulates a joint optimization problem aimed at minimizing the "bubble rate." By leveraging an Alternating Optimization (AO) algorithm, it jointly optimizes the model split point, micro-batch number, batch size, and time slot allocation to mitigate the synchronization latency caused by device heterogeneity.
Different from the aforementioned pipeline strategies, Hourglass~\cite{he2025hourglass} targets the system scalability bottlenecks inherent in server-side resource management. It proposes a resource-multiplexed data parallelism architecture that decouples the maintenance of server-side model partitions from the massive number of clients. By mapping clients to a limited pool of GPU-resident models, Hourglass effectively eliminates the significant computation and storage overheads caused by frequent model context switching.

\begin{table*}[t]
\setlength{\tabcolsep}{2pt}
\centering

\caption{Global joint optimization methods. The ``Sys. Hetero.'' column denotes consideration of system heterogeneity across clients. A checkmark ($\checkmark$) indicates the variable is optimized. Comm. Res.: Communication Resources; Comp. Res.: Computation Resources.}
\vskip -5pt
\label{tab:optimization}
\resizebox{\textwidth}{!}{%
\begin{tabular}{l c c c p{3.5cm} c ccccc p{4.0cm}}
\toprule

\multirow{2}{*}{\textbf{Method}} & \multirow{2}{*}{\textbf{Year}} & \multirow{2}{*}{\textbf{Split Mode}} & \multirow{2}{*}{\textbf{Framework}} & \multirow{2}{*}{\textbf{Optimization Objective}} & \multirow{2}{*}{\textbf{Sys. Hetero.}} & \multicolumn{5}{c}{\textbf{Decision Variables}} & \multirow{2}{*}{\textbf{Solution Method}} \\
\cmidrule(lr){7-11}
 & & & & & & Cut Layer & Batch Size & Update Freq. & Comm. Res. & Comp. Res. & \\
\midrule

CPSL~\cite{wu2023split} & 2023 & Two-part & GSFL & Training Latency & \checkmark & \checkmark & & & \checkmark & & Two-timescale (SAA + Gibbs Sampling)\\
U-PSL~\cite{lyu2023optimal} & 2023 & Three-part & PSL & Training Latency & & \checkmark & & & & \checkmark & LSCRA(Exhaustive + Bisection) \\
\midrule

JSFL~\cite{xu2023accelerating} & 2024 & Two-part & SFL & Training Latency & & \checkmark & & & \checkmark & & Alternating Optimization \\
EPSL~\cite{lin2024efficient} & 2024 & Two-part & PSL & Training Latency & & \checkmark & & & \checkmark & & Block Coordinate Descent \\
AdaSFL~\cite{liao2023accelerating} & 2024 & Two-part & SFL & Training Latency & \checkmark & & \checkmark & \checkmark & & & Linear Programming \\
MergeSFL~\cite{liao2024mergesfl} & 2024 & Two-part & SFL & Sync. Wait Time & \checkmark & & \checkmark & & & & Genetic Algorithm \\
EdgeSplit~\cite{zhang2024resource} & 2024 & Two-part & SFL & Training Latency & \checkmark & \checkmark & & & \checkmark & & Alternating Algorithm \\
ASL~\cite{li2024adaptive} & 2024 & Two-part & SL & Training Latency + Energy Constraints & & \checkmark & & & & \checkmark & Lyapunov Optimization + Two-layer Optimization \\
\midrule

\llm{PASFL$^*$~\cite{chen2025pasfl}} & 2025 & Two-part & SFL & Latency + Server Energy + Privacy & \checkmark & \checkmark & & & \checkmark & & BCD-based Iterative Algorithm\\
SAFSL~\cite{ao2025semi} & 2025 & Two-part & SFL & Training Latency & \checkmark & \checkmark & & & \checkmark & \checkmark & Lyapunov (Online Drift-Plus-Penalty) \\
\llm{SFL-LLM$^*$~\cite{zhao2025efficient}} & 2025 & Two-part & SFL & Training Latency & \checkmark & \checkmark & & & \checkmark & & BCD-based Iterative Algorithm \\
\llm{CARD$^*$~\cite{li2025energy}} & 2025 & Two-part & SL & Latency + Server Energy & & \checkmark & & & & \checkmark & Problem Decomposition / Convex Optimization \\
AdaptSFL~\cite{lin2025adaptsfl} & 2025 & Two-part & SFL & Training Latency & \checkmark & \checkmark & & \checkmark & & & Block Coordinate Descent \\
SFL-GA~\cite{chen2021communication} & 2025 & Two-part & SFL & Latency + Convergence & \checkmark & \checkmark & & & \checkmark & \checkmark & DDQN + CVX \\
HASFL~\cite{lin2025hasfl} & 2025 & Two-part & SFL & Training Latency & \checkmark & \checkmark & \checkmark & & & & BCD-based Iterative Algorithm \\
HeteroSFL~\cite{sun2025split} & 2025 & Two-part & SFL & Training Latency & \checkmark & \checkmark & & & \checkmark & \checkmark & SAA + GA; Lagrangian + B\&B \\
\llm{SFT-LLM$^*$~\cite{zhang2025split}} & 2025 & Two-part & SFL & Training Latency & & \checkmark & & & \checkmark & & Decomposition + Augmented Lagrangian + SQP \\

\bottomrule
\end{tabular}%
}
\vskip -10pt
\end{table*}

\subsection{Global Joint Optimization}

In the context of Split-based LLM-FT, achieving optimal training performance necessitates the intricate orchestration of heterogeneous resources, including client-server computational power and dynamic communication bandwidth. This process is governed by deeply coupled decision variables. For instance, the selection of the model split point dictates the distribution of computational loads and local memory consumption, which in turn acts as a strict constraint on the feasible batch size, a parameter that directly governs transmission throughput. Consequently, optimizing any single variable in isolation fails to resolve the systemic trade-offs between local resource constraints and global training efficiency. To address this, global joint optimization formulates the fine-tuning process as a constrained optimization problem, jointly adjusting parameters such as the cut layer, batch size, and update frequency to minimize objectives like training latency and energy consumption, as comprehensively summarized in Table \ref{tab:optimization}.

\section{Privacy and Security in Split-based LLM-FT}
\label{section: privacy}
\begin{figure}[t]
    \centering
    \includegraphics[width=1\linewidth]{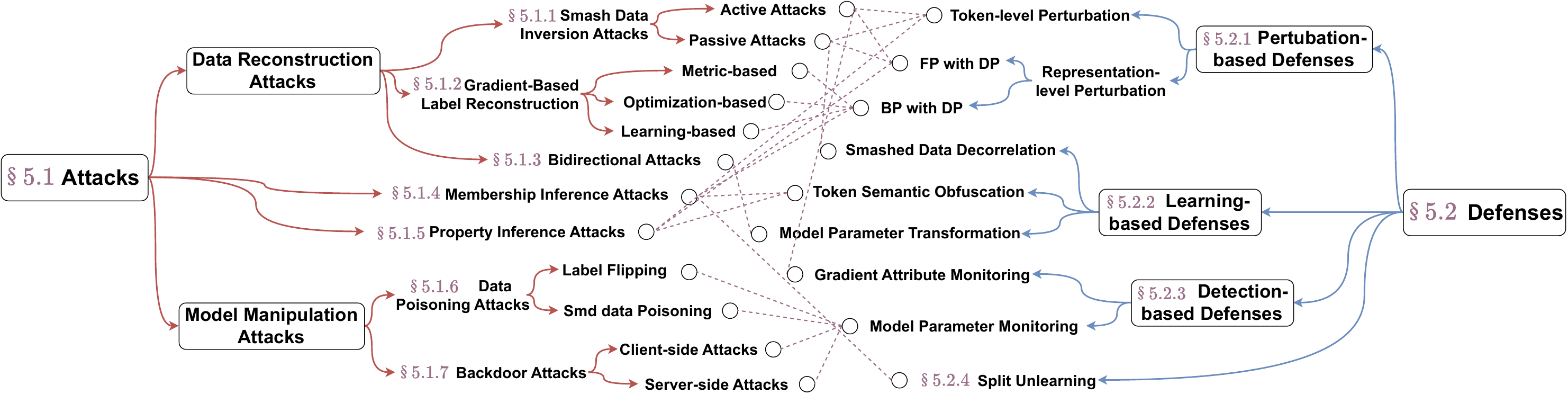}
    \vskip -5pt
    \caption{Taxonomy of attacks and defenses in split-based LLM-FT.}
    \vskip -10pt
    \label{fig:attack_defense_taxonomy}
\end{figure}

Despite its architectural advantages for resource-constrained clients and inherent data locality, Split Learning (SL) remains susceptible to various security and privacy risks. While the SL paradigm mitigates the direct exposure of raw inputs and labels by transmitting only intermediate activations—commonly referred to as smashed data—and gradients, these variables still carry significant sensitive features that can be exploited. Recent studies have demonstrated that malicious actors can leverage these intermediate representations to reconstruct private information through methodologies such as model inversion and label inference attacks, or otherwise compromise the integrity of the training process. Accordingly, this section provides a systematic overview of the security landscape in split-based LLM-FT by first categorizing and introducing prominent attack methodologies targeting SL architectures, with a specific focus on those applicable to Large Language Models, and subsequently summarizing the existing defense strategies designed to mitigate these threats and protect the privacy of collaborative fine-tuning. Fig.~\ref{fig:attack_defense_taxonomy}. presents a systematic taxonomy of these attacks and defenses in SL. 

\subsection{Attacks on Split-based LLM-FT}
The security vulnerabilities in Split-based LLM-FT primarily arise from the exchange of intermediate activations, gradients, and model states, which inadvertently expose a broad attack surface, as these high-dimensional representations often retain potent semantic features. Due to the inherent nature of Transformer architectures, such as residual connections that facilitate the preservation of input information throughout deep layers, malicious actors can reconstruct sensitive information including original text sequences and proprietary labels without direct access to raw data. This subsection categorizes the prevailing attack vectors and introduces their underlying technical mechanisms. 
\lzh{Notably, methods marked with an asterisk (*) represent attack and defense mechanisms designed for or tested on LLMs. Unmarked methods represent conventional security approaches whose core insights are theoretically transferable to LLM.}

\subsubsection{Smashed data inversion attacks}
\label{subsubsection:smdia}
During forward propagation, attackers can leverage transmitted smashed data to reconstruct original inputs through model inversion attacks. These threats are categorized into two types based on the adversary’s adherence to the training protocol: (1) active attacks, in which the adversary deliberately manipulates the training procedure or violates established protocols to facilitate more effective information extraction; and (2) passive attacks, where the attacker strictly adheres to the SL protocol by honestly computing and returning gradients without tampering with the training process or model update logic.

\noindent \textbf{Active attacks.} In an active attack, a malicious server hijacks or tampers with the gradients backpropagated to the client. The objective is to intentionally guide the client-side model toward specific behaviors or feature representations that expose sensitive information, thereby significantly enhancing the server's ability to reconstruct private data. 
A prominent example of an active attack is the Feature-Space Hijacking Attack (FSHA)~\cite{gawron2022feature,pasquini2021unleashing}. In this scheme, a malicious server hijacks the training process by providing forged gradients generated from an internal discriminator. These adversarial gradients intentionally guide the client-side model to map its private data into a target, invertible feature space. Once this alignment is achieved, the server utilizes a locally pre-trained decoder to reconstruct the original inputs from the transmitted smashed data with high fidelity. This attack effectively nullifies the privacy-preserving properties of the split learning framework by forcing the client model into an insecure functional state. Despite its potency, FSHA inherently decouples the training process from the intended main task, leaving it vulnerable to detection mechanisms that probe gradient sensitivity toward mislabeled or "fake" samples. To mitigate this vulnerability, recent advancements focus on enhancing the stealth and task-alignment of hijacking attacks. SIA~\cite{yu2024sia}, for instance, integrates a server-side main task model that simultaneously disguises the hijacking process and supervises the client-side model. This dual-purpose architecture ensures the extraction of semantic-rich features, rendering the attack far less conspicuous than methods relying solely on adversarial loss.Building on this stealth-oriented logic, SplitSpy~\cite{fu2023focusing} addresses the threat of gradient scrutiny by employing a "Legitimate Shadow Model" to pinpoint "fake" samples via their characteristically high prediction errors. By designating the top $\lambda$ percent of high-loss samples as decoys, SplitSpy implements a differential gradient strategy: it returns legitimate gradients for detected decoys to mimic honest behavior while surreptitiously injecting malicious hijacking signals into real samples. This selective execution allows the adversary to achieve high-fidelity data reconstruction while effectively bypassing sophisticated detection protocols.

\noindent \textbf{Passive attacks.} In contrast, passive attacks operate under the honest-but-curious threat model. In this paradigm, the adversary attempts to infer private training instances solely from the received smashed data while strictly adhering to the split learning protocol. This adherence requires the faithful calculation and transmission of gradients without tampering with the training process or the model update logic. Because these attacks do not disrupt the primary training objective, they exhibit high stealthiness and can effectively circumvent active defense mechanisms designed to identify gradient anomalies. Existing literature typically classifies passive inversion strategies into two distinct paradigms: \textit{optimization-based} and \textit{learning-based} approaches~\cite{he2019model}.

\noindent\textit{(1) Optimization-based approaches}~\cite{erdougan2022unsplit, li2023gan}.
These methods model data reconstruction as an iterative optimization problem, aiming to find an input that matches the observed smashed data or gradients.
Specifically, UnSplit~\cite{erdougan2022unsplit} models the attack as an optimization problem and solves it using coordinate gradient descent. The attacker alternately updates the parameters of the reconstructed input and the proxy model. This method allows the server to simultaneously steal the client's model functionality without accessing auxiliary public datasets or query functions, and recovers the private input with high fidelity by minimizing mean squared error and total variation loss. Similarly, GLASS~\cite{li2023gan} reformulates the attack as a latent space search within a pre-trained StyleGAN. It reconstructs the private input by optimizing latent codes to minimize the distance between the intermediate features of the generated and target data. This two stage strategy explores the entangled space to escape local minima and refines the result in the disentangled space. To address optimization challenges at deeper layers, GLASS++~\cite{li2023gan} improves this method by employing an encoder to map features directly to the latent space. This provides a robust initialization for the latent search, preventing the optimization from getting trapped in local minima and ensuring stability when spatial information is limited.

\noindent\textit{(2) Learning-based approaches}~\cite{gao2023pcat,zhang2024functionality,zeng2025gan,luo2023fsa,xu2024stealthy,zhu2025passive}.
Unlike optimization methods, learning-based approaches typically involve constructing a shadow model or pseudo-client trained on an auxiliary dataset to learn an inverse mapping from smashed data back to the input space.
PCAT~\cite{gao2023pcat} trains a pseudo-client model on a small auxiliary dataset by minimizing the downstream task loss (e.g., Cross-Entropy) calculated via the server model. Based on this simulation, the server trains a reverse mapping network to reconstruct original inputs from intercepted smashed data. Furthermore, PCAT adopts a ''stealing while training'' strategy that compels the pseudo-client to track the real client's learning trajectory. This prevents convergence to local optima with mismatched internal features, significantly enhancing both functionality stealing and data reconstruction.
However, when victim clients employ label transformation defenses (e.g., mapping real labels to fake or encoded ones), Basic PCAT fails as the server's top model learns incorrect mappings, misleading the pseudo-client. Improved PCAT~\cite{zhang2024functionality} addresses this by introducing a pseudo-top Model to substitute the compromised real top layer, effectively eliminating dependency on the server's original task objective. By leveraging the shared intermediate model for feature extraction and employing a multi-model parallel training strategy to identify the best match, Improved PCAT successfully steals authentic functionality despite label tampering, significantly enhancing attack robustness.
MAERA~\cite{zeng2025gan} contends that relying exclusively on task-oriented objectives (e.g., cross-entropy) is insufficient for high-fidelity data reconstruction. While such methods may achieve functional equivalence, they suffer from a significant ``feature space gap,'' yielding reconstructed data lacking original semantic details. To bridge this disparity, MAERA proposes a GAN-based framework employing a hybrid loss function. Specifically, the discriminator minimizes the divergence between shadow and authentic features, ensuring the shadow model's output statistically aligns with the real client's latent space. 
Other approaches adopt similar principles by incorporating adversarial losses during the reconstruction phase to enhance the fidelity of the recovered data. For instance, FSA~\cite{luo2023fsa} employs a cycle-consistent generative adversarial network to align smashed data with a feature space defined by a pre-trained autoencoder. Similarly, FORA~\cite{xu2024stealthy} leverages domain discriminator loss combined with multi-kernel maximum mean discrepancy to project and align the output features of the pseudo-client and the real client within a shared latent space.
SDAR~\cite{zhu2025passive} refines this using dual adversarial discriminators to respectively align the simulator's intermediate representations with authentic client outputs, and the decoder's reconstructions with real natural sample distributions.

\begin{table*}[t]
\centering
\caption{Summary of smashed data inversion attacks on Split-based LLM-FT. ($\checkmark$: required/known, $\times$: not required/unknown)}
\vskip -5pt
\label{tab:attack_summary}
\resizebox{\textwidth}{!}{%
\begin{tabular}{l c c c c c c}
\toprule
\multirow{2}{*}{\textbf{Method}} & \multirow{2}{*}{\textbf{Attack Type}} & \multirow{2}{*}{\textbf{Reconstruction Paradigm}} & \multirow{2}{*}{\textbf{Attack Phase}} & \multicolumn{3}{c}{\textbf{Attacker's Capability}} \\
\cmidrule(lr){5-7}
 & & & & \textbf{Client Arch.} & \textbf{Client Weights} & \textbf{Aux. Dataset} \\
\midrule
\textbf{FSHA}~\cite{gawron2022feature} & Active & Learning-based & Training & $\times$ & $\times$ & $\checkmark$ \\
\textbf{SIA}~\cite{yu2024sia} & Active & Learning-based & Training & $\times$ & $\times$ & $\checkmark$ \\
\textbf{SplitSpy}~\cite{fu2023focusing} & Active & Learning-based & Training & $\times$ & $\times$ & $\checkmark$ \\
\midrule
\textbf{UnSplit}~\cite{erdougan2022unsplit} & Passive & Optimization-based & Inference & $\checkmark$ & $\times$ & $\times$ \\
\textbf{GLASS}~\cite{li2023gan} & Passive & Optimization-based & Inference & $\checkmark$ & $\checkmark$ & $\checkmark$ \\
\textbf{GLASS++}~\cite{li2023gan} & Passive & Hybrid & Inference & $\checkmark$ & $\checkmark$ & $\checkmark$ \\
\textbf{PCAT}~\cite{gao2023pcat} & Passive & Learning-based & Training & $\times$ & $\times$ & $\checkmark$ \\
\textbf{Improved PCAT}~\cite{zhang2024functionality} & Passive & Learning-based & Training & $\times$ & $\times$ & $\checkmark$ \\
\textbf{MAERA}~\cite{zeng2025gan} & Passive & Learning-based & After convergence & $\times$ & $\times$ & $\checkmark$ \\
\textbf{FSA}~\cite{luo2023fsa} & Passive & Learning-based & Inference & $\times$ & $\times$ & $\checkmark$ \\
\textbf{FORA}~\cite{xu2024stealthy} & Passive & Learning-based & Training & $\times$ & $\times$ & $\checkmark$ \\
\textbf{SDAR}~\cite{zhu2025passive} & Passive & Learning-based & Training & $\times$ & $\times$ & $\checkmark$ \\
\bottomrule
\end{tabular}%
}
\vskip -12pt
\end{table*}

\subsubsection{Gradient-based label reconstruction}
In u-shaped SL, even though clients keep labels locally, the gradients sent to the server still carry hidden label information. This occurs because gradients are calculated based on the difference between the model's prediction and the actual label. As a result, a malicious server can analyze these gradients to recover private labels.
The risk of such label leakage is further exacerbated in the context of LLMs. Specifically, for Causal Language Modeling (CLM) tasks that utilize autoregressive mechanisms, the supervision signal (i.e., the label) is essentially the input sequence shifted by one token. This implies that in such scenarios, stealing labels is effectively equivalent to reconstructing the private raw input text~\cite{liu2025dualguard,chen2024unveiling}.
Existing gradient-based label reconstruction attacks can be categorized into three streams based on their underlying mechanisms: metric-based gradient analysis, optimization-based gradient matching, and learning-based label inversion.
  
\noindent \textbf{Metric-based gradient analysis attacks.}
These attacks exploit the mathematical properties of loss functions. By calculating specific statistical metrics (e.g., gradient norm, distance, similarity) on the gradients or smashed data at the split layer, attackers can directly distinguish ground-truth labels.
Li et al.~\cite{li2021label} recover private labels by analyzing returning gradients via two methods. The norm-based attack exploits asymmetrical gradient norm distributions, where positive samples generate larger gradients, allowing threshold-based recovery. Meanwhile, the Direction-based Attack leverages the geometric property that same-class gradients form acute angles, enabling reconstruction through cosine similarity and clustering.
However, these methods are specifically designed for binary classification tasks and may not be directly applicable to more complex multi-class scenarios.
Liu et al.~\cite{liu2023distance} address this with a distance-based attack unifying gradient and smashed data metrics into euclidean distance. Furthermore, they utilize pre-trained models to extract auxiliary baseline features, inferring labels via target-auxiliary distances.
In a significant extension, Liu et al.~\cite{liu2024similarity} exploit latent grouping characteristics by integrating K-means clustering with the K-M matching algorithm. This method maps latent data clusters to ground-truth labels using limited auxiliary samples, demonstrating superior scalability on complex datasets and enhanced robustness against gradient compression.

\noindent \textbf{Optimization-based gradient matching attacks.}
This category of attacks models label reconstruction as an iterative optimization task based on gradient matching. The adversary initializes dummy labels and a surrogate model to generate virtual gradients, then iteratively optimizes them by minimizing the discrepancy with the client's real gradients to recover the true labels.
UnSplit~\cite{erdougan2022unsplit} executes label inference by minimizing the Mean Squared Error (MSE) between the client's actual gradients and a surrogate's proxy gradients across all label candidates. Identifying the closest gradient profile enables perfect label inference. However, this attack is primarily effective when the label owner's local model is shallow (e.g., one layer deep).
To overcome this limitation, ExPLoit~\cite{kariyappa2023exploit} incorporates label prior and label entropy regularization into its loss function. By enforcing alignment between surrogate predictions and labels, this method avoids poor local minima, maintaining efficacy even against multi-layer networks.
Xie et al.~\cite{xie2023label} demonstrated that by utilizing a training accuracy loss, label inference attacks in split learning can be extended from classification problems to regression problems.
SplitAUM~\cite{zhao2024splitaum} introduces a semi-supervised clustering strategy (e.g., K-Means) to initialize dummy labels with feature distributions matching the ground-truth. This addresses the limitation of random initialization, which often misleads the model.
Leveraging the "not-too-far" characteristic of LLM fine-tuning, \llm{Chen et al.$^*$~\cite{chen2024unveiling}} proposed directly utilizing the tail layers of the pre-trained model as the surrogate model. By employing the pre-trained weights as a robust prior, this strategy creates a semi-white-box environment that eliminates the need to optimize a randomly initialized surrogate from scratch. Crucially, this setting empowers the adversary to effectively deploy existing Gradient Matching Attacks (GMA), including DLG~\cite{zhu2019deep}, iDLG~\cite{zhao2020idlg}, TAG$^*$~\cite{deng2021tag}, and LAMP$^*$~\cite{balunovic2022lamp}, to recover private labels. Consequently, the approach significantly enhances attack accuracy by combining the stability of pre-trained priors with the optimization power of established gradient inversion algorithms.

\noindent \textbf{Learning-based label inversion attacks.}
Unlike optimization-based methods that rely on per-sample iteration, learning-based attacks aim to functionally complete the split network. The adversary leverages auxiliary data to train a pseudo-tail model that acts as a substitute for the victim's original tail model, thereby restoring the direct mapping from smashed data to private labels. Once this model completion process is finished, the adversary can seamlessly append the pseudo-tail model to the split layer for efficient, end-to-end label inference.
Fu et al.~\cite{fu2022label} proposed a passive inference attack based on model completion, where the adversary appends an inference head to their trained bottom model and fine-tunes it using limited auxiliary labeled data via semi-supervised learning. Once fine-tuned, this completed model is used to directly predict the labels of any target sample based solely on the adversary's local features. They further enhanced this approach with an active attack using a malicious local optimizer, which adaptively scales gradients to induce the global model to rely more heavily on the adversary's features, thereby improving the expressiveness of the bottom model for label inference. 
To address the overfitting issue where pseudo-tail models memorize auxiliary data instead of learning transferable features, SDAR~\cite{zhu2025passive} employs random label flipping. This regularization strategy effectively enhances the model's generalization ability, enabling accurate label inference in u-shaped SL.

\subsubsection{Bidirectional reconstruction attacks}
Previous attacks typically exploit a single source of information, utilizing either the smashed data from forward propagation or the gradients from backward propagation to recover private labels or input text. However, in the context of Split-based LLM fine-tuning, these two data streams contain complementary semantic information. Furthermore, due to the autoregressive nature of Causal Language Modeling (CLM), the input text and the supervision labels, which represent the next-token prediction, are semantically equivalent as they differ by only a single token shift. Bidirectional reconstruction attacks aim to integrate the leaked information from both forward and backward propagation to achieve higher fidelity in data recovery. This approach effectively overcomes the inherent information bottlenecks associated with unidirectional attacks.
\llm{Chen et al.$^*$~\cite{chen2024unveiling}} pioneered this direction by proposing Bidirectional Semi-white-box Reconstruction (BiSR), a framework specifically tailored for Split-based private LLM fine-tuning. Grounded in the "not-too-far" assumption, which posits that the parameters of a fine-tuned model remain spatially close to those of the publicly available pre-trained model, BiSR employs a synergistic two-pronged strategy. 
It integrates the Semi-white-box Inversion Paradigm (SIP), where the adversary leverages pre-trained weights to train an inverse network mapping smashed data back to input embeddings, with a backward gradient matching mechanism to extract token information. BiSR's critical innovation lies in deeply coupling these processes: high-level semantic features from the forward SIP module regularize the backward gradient matching objective. This semantic guidance prevents gradient matching from falling into poor local minima, reconstructing private fine-tuning data with significantly higher accuracy and coherence than unidirectional leakage methods.

\subsubsection{Membership inference attacks}
Membership Inference Attacks (MIA) are designed to infer whether a specific data record was included in the training dataset of a target model. Traditional MIA operates in a black-box setting, requiring query access to the target model. Attackers train shadow models using the same ML platform or a similar architecture to replicate the target's behavior and infer membership. However, in Split Learning, attackers lack access to the client-side model segment, rendering direct queries impossible. To address this, Chen et al.~\cite{chen2020practical} proposed a novel "Transfer-Inherit" strategy. They substitute the inaccessible client layers with a pre-trained feature extractor and splice it with the accessible server-side layers. By fine-tuning this feature extractor on a shadow dataset, the attacker can effectively mimic the target model's behavior with limited data. This local shadow model is then used to train the attack model, enabling membership inference without querying the actual target system, thus achieving a "query-free" attack. To address the significant prediction misalignment when using shadow model responses to estimate the target model's memory, Yi et al.~\cite{yi2025membership} proposed the Knowledge Transfer Membership Inference Attack (KTMIA), which reconstructs both models by splicing pseudo-clients onto frozen server parameters to ensure aligned membership signals.

\subsubsection{Property inference attacks}
Property Inference Attacks~\cite{pasquini2021unleashing, mao2023secure} target the extraction of specific sensitive attributes that are often orthogonal to the primary learning task, such as author identity, political inclination, or underlying emotional sentiments. Compared to full data reconstruction, these attacks typically exhibit lower implementation complexity and greater robustness against existing defense mechanisms. In the context of split learning, an attacker executes this by replacing the original server-side model with a specific property classifier. By leveraging a labeled auxiliary dataset for supervised training, the attacker effectively coerces the client model to align its feature space with representations that explicitly encode the target attribute, thereby enabling the precise inference of private information.


\subsubsection{Data poisoning attacks}
Data poisoning attacks involve the injection of corrupted or malicious samples into the training dataset during the model's training phase, thereby compromising its integrity or usability. In the context of multi-client Split-based LLM-FT, such as multi-client sequential SL or SFL, one or more malicious clients can manipulate their local training data to significantly degrade the performance of the global model.

\noindent \textbf{Label flipping attacks.} 
Label-flipping attack in Split-based LLM-FT is a form of dirty-label poisoning where malicious clients intentionally manipulate their local training labels to induce erroneous model predictions and degrade the system's overall robustness.
Kohankhaki et al.~\cite{kohankhaki2023detecting} explored label-poisoning threats within multi-client sequential SL, revealing that the integrity of global models is highly sensitive to even minor data manipulation by malicious participants.
Gajbhiye et al.~\cite{gajbhiye2022dpa} analyzed the vulnerability of SFL variants (SFLV1 and SFLV2) to label flipping attacks. While both variants proved susceptible, SFLV2 demonstrated significantly lower robustness. The study also concluded that untargeted attacks cause more severe degradation to the global model than targeted attacks.
Ismail and Shukla~\cite{ismail2023analyzing} further expanded the threat landscape of SFL by introducing a novel distance-based poisoning strategy. Unlike previous works that relied on random or fixed label selection, this approach optimizes the attack by selecting target labels that maximize the Euclidean distance between input samples. Their research demonstrated that this distance-oriented method achieves a more potent degradation of model performance than traditional targeted attacks.

\noindent \textbf{Smash data poisoning attacks.} 
While label flipping attacks target the supervision signal directly, they are often directly adapted from Federated Learning (FL) without exploiting the unique architectural characteristics of split-based learning. 
Wu et al.~\cite{wu2024evaluating} proposed a novel attack specifically tailored for SFL, named Smash Poisoning, which applies gradient ascent perturbations to the smashed data by leveraging historical gradients. Furthermore, combining this method with Label Poisoning was shown to significantly enhance the attack's potency.
Recently, a hybrid attack strategy named MISA~\cite{wan2024misa} was proposed to induce global model misalignment by simultaneously targeting the server-side top model and the client-side model. This method indirectly poisons the top model through statistically crafted perturbations on smashed data, while employing Thompson Sampling and a Search-and-Locate (SnL) mechanism to optimize the bottom model manipulation as a Min-Sum problem.

\subsubsection{Backdoor attacks}
Backdoor attacks~\cite{gu2017badnets} refer to an adversary implanting a specific pattern or "trigger" into a machine learning model. This allows the trained model to maintain state-of-the-art performance on clean validation samples to remain stealthy while being subtly manipulated to output a specific target class whenever the trigger appears in the input during testing. Traditional backdoor attacks are typically implemented through training set poisoning where a portion of the training data is modified to include the trigger and reassigned to the attacker's target label. In the context of Split-based LLM-FT, these attacks can be categorized into client-side backdoor attacks and server-side backdoor attacks, depending on which party initiates the attack.

\noindent \textbf{Client-side backdoor attacks.} 
In scenarios where the client possesses both raw inputs and ground-truth labels, it can readily deploy traditional backdoor strategies, such as injecting specific triggers into training samples and flipping their labels to implant malicious associations~\cite{yu2023backdoor}. However, in the context of two-party SL, where data and labels are vertically partitioned, the client holds only the input data. This lack of access to supervision signals makes executing backdoor attacks significantly more challenging, as the adversary cannot directly manipulate the label to establish the trigger-target correlation.
Yu et al.~\cite{yu2023backdoor} propose that attackers can employ label inference attacks to bypass feature-label isolation and identify target samples. Furthermore, by integrating an auxiliary model at the client side to distinguish between clean and backdoor features, the attacker forces the client model to map triggered samples into a distinct, isolated region of the feature space, thereby significantly enhancing the success rate of the backdoor attack.
He et al.~\cite{he2023backdoor} leverage the Mean-shift algorithm to extract a high-density "trigger vector" from the local embeddings of a small auxiliary set belonging to the target class. By consistently injecting this vector during training, the malicious client forces the top model to establish a strong associative mapping between the trigger and the target label. During inference, the attacker can hijack the classification of any sample by replacing its local embedding with this trigger, effectively overriding host-side features to trigger the desired misclassification.
Bai et al.~\cite{bai2023villain} propose VILLAIN, a framework utilizing gradient-based label inference. By swapping sample embeddings with a target seed and monitoring gradient magnitude, the attacker identifies target labels. To enhance stealth, a semi-supervised classifier restricts this inference to selected candidates. VILLAIN then injects additive triggers into high-variance features, employing trigger randomization and learning rate adjustment to ensure robust backdoor implantation.
PPoison~\cite{liu2026ppoison} injects a backdoor by introducing a pluggable trigger layer that directly transforms the smashed data output by the local model, optimized via a local surrogate classifier. In normal mode, the trigger layer remains inactive, transmitting original features to ensure accurate predictions; in attack mode, the layer is activated to convert embeddings from arbitrary inputs into poisoned representations, compelling the server-side model to misclassify them as a preset target label.

\noindent \textbf{Server-side backdoor attacks.} 
In the server-side attack model, due to the honesty of the clients, attackers can neither access the raw input nor inject triggers, leading to significant structural barriers for backdoor attacks.
Tajalli et al.~\cite{tajalli2023feasibility} proposed the first threat model for server-side backdoor attacks in SL. The authors introduced two methods, the surrogate client and the injector autoencoder, to inject backdoors without access to client inputs. However, their findings indicate that these attacks are only viable on simple datasets and largely fail as data complexity increases.
Targeting scenarios where the server holds the labels, IBAS leverages a small amount of public data to train a ''parrot model'' that approximates the client's feature space. During the forward pass of training, the server replaces a subset of the client's uploaded intermediate activations with poisoned ones generated by the parrot model. By modifying the corresponding labels and backpropagating the gradients, IBAS~\cite{xi2025ibas} effectively guides the client model to embed a backdoor while preserving the performance of the main task.
In scenarios where the server does not possess ground truth labels, executing server-side attacks becomes significantly more challenging
Yu et al.~\cite{yu2023backdoor} leveraged a shadow model trained on domain-consistent data to pre-simulate backdoor injection. By employing a GAN-based discriminator, they forced the alignment of the client’s feature space with that of the shadow model, thereby endowing the client with indirect backdoor extraction capabilities. Additionally, an auxiliary model maximizes the separability of backdoor features to prevent them from being diluted during training.
However, to mitigate the dependency on large-scale public datasets and improve main task alignment, SFI~\cite{yu2024chronic} further deconstructs the attack into three distinct stages (Steal, Finetune, and Implant). This staged design allows the model to first learn the feature space before injecting the backdoor, exerting finer control over the attack process. This approach ensures stability and effectiveness, thereby improving the attack's accuracy in complex scenarios.
Dullahan~\cite{pu2024dullahan} is a stealthy post-training attack strategy designed for the without-label-sharing architecture, diverging from previous methods by refraining from manipulating intermediate parameters (e.g., gradients) during the collaborative training phase. To overcome the lack of label access, the method constructs a surrogate model to approximate the client's feature output and employs K-means clustering to identify a "target embedding anchor" as a substitute for the inaccessible target label. The backdoor is subsequently injected by fine-tuning the server network to map selected trigger embeddings to 
this anchor, thereby achieving high stealthiness against gradient-based detection mechanisms.

\subsection{Defenses on Split-based LLM-FT}
To mitigate the security and privacy risks inherent in the collaborative training process, research has increasingly focused on developing robust defense mechanisms for Split-based LLM-FT. The primary objective of these strategies is to obfuscate or protect the sensitive features embedded within intermediate representations and gradients, thereby preventing adversaries from successfully executing inversion or inference attacks. However, designing effective defenses requires navigating a complex trade-off between the level of privacy protection, system execution efficiency, and the final model performance. This subsection categorizes the prevailing defense strategies into three primary paradigms based on their underlying protection mechanisms: perturbation-based, learning-based, and cryptographic-based defenses.

\subsubsection{Perturbation-based defenses} 
Perturbation-based defenses involve introducing controlled randomness into the training pipeline to obfuscate sensitive information while maintaining as much model utility as possible. These methods are frequently grounded in the principles of Differential Privacy (DP)~\cite{dwork2006calibrating}, providing a quantifiable and mathematically rigorous trade-off between privacy protection and fine-tuning accuracy. In the context of Split-based LLM-FT, these strategies can be broadly categorized based on the granularity of noise injection: token-level perturbation, which operates on the discrete input sequence, and representation-level perturbation, which injects noise into continuous vectors such as embeddings, activations, and gradients.

\noindent \textbf{Token-level perturbation (discrete perturbation).} 
Token-level Perturbation targets the discrete input space of LLMs by substituting sensitive tokens within the input sequence with generic or alternative replacements. By perturbing the sequence while preserving essential semantic structures, this strategy mitigates the risk of private text reconstruction from the earliest stage of the training pipeline.
Embedding d$\chi$-Privacy (d$\chi$P)~\cite{chatzikokolakis2013dxp} injects Laplacian noise into embedding outputs and maps the perturbed results to their nearest valid neighbors in the embedding space to preserve semantic coherence while providing privacy protection. 
To mitigate utility degradation from uniform noise, \llm{SAP$^*$~\cite{shen2025sap}} employs Contributing-Token-Identification (CTI) to identify task-critical tokens and adaptively allocate privacy budgets by reducing perturbations on important tokens, thereby achieving a superior privacy-utility trade-off.
\llm{SanText$^*$~\cite{yue2021santext}} generates natural, high-utility text satisfying local differential privacy by using probability sampling based on word embedding Euclidean distances to replace original tokens with semantically similar substitutes.
However, SanText uses the entire vocabulary as its sampling candidate set, which can easily introduce semantically irrelevant interference and lead to redundant privacy costs when privacy budgets are tight. In contrast, \llm{CusText$^*$~\cite{chen2023custext}} effectively eliminates irrelevant terms by allocating a smaller, semantically highly relevant customized output set to each token, significantly improving the utility of anonymized data while reducing privacy overhead. Furthermore, \llm{Rantext$^*$~\cite{tong2025inferdpt}} introduces a random adjacency list mechanism that leverages the Laplace distribution to dynamically determine the list size for each token, thereby increasing the difficulty for adversaries to reconstruct the raw tokens. \llm{Cape$^*$~\cite{wu2025cape}} integrates a hybrid utility function combining Euclidean token distances with client-side contextual logits and employs a bucketized sampling mechanism to mitigate the large-vocabulary long-tail phenomenon, effectively suppressing low-utility candidate probabilities to enhance semantic coherence and quality while ensuring differential privacy.

\noindent \textbf{Representation-level perturbation (continuous noise injection).} 
This approach involves injecting continuous noise such as Gaussian or Laplacian noise into high-dimensional intermediate values including embedding vectors, smashed data (activations), and backpropagated gradients.

\noindent\textit{(1) Forward Propagation with DP Noise.}~\cite{pan2020privacy, mai2023snd, du2023dpforward, wang2024selective, pham2024enhancing, huang2025sldp}
This strategy injects noise into intermediate values including embedding vectors and smashed data during forward propagation before they are transmitted from the client to the server. 
Pan et al.~\cite{pan2020privacy} mitigate information leakage by injecting Laplacian noise into embedding vectors. 
To overcome the utility degradation inherent in traditional perturbation mechanisms, \llm{SnD$^*$~\cite{mai2023snd}} implements $d_{\chi}$-privacy while deploying a pre-trained denoising module on the client side. This architecture enables localized correction of the perturbed inference results, thereby maximizing the preservation of model utility.
\llm{DP-forward~$^*$\cite{du2023dpforward}} schemes incorporate noise into activation values. These privacy-preserving techniques are designed to prevent adversaries from inferring original inputs from perturbed smashed data, thereby ensuring robust privacy protection throughout the forward propagation process.
To address the utility degradation caused by uniform token perturbation in DP-forward, \llm{Wang et al.$^*$~\cite{wang2024selective}} propose SLDP-FT. This framework selectively injects noise only into identified sensitive tokens. Furthermore, it dynamically calibrates the noise scale across hidden layers using a privacy forward weight, achieving a superior privacy-utility trade-off.

\noindent\textit{(2) Backward propagation with DP noise.}~\cite{yang2022differentially,wu2023federated,qiu2023evaluating,yang2022differentially}
This strategy injects noise during the backward pass to protect sensitive information within transmitted gradients. It involves adding Gaussian or Laplacian noise to model gradients, cut layer gradients, and labels to effectively mitigate the risk of label reconstruction attacks. 
Gawron et al.~\cite{gawron2022feature} evaluated the robustness of DP-SGD-based client defense in split learning by applying gradient clipping and Gaussian noise to client-side model to suppress FSHA reconstruction accuracy and delay privacy leakage.
GradPerturb~\cite{yang2022differentially} provides provable differential privacy by injecting directional noise at the final hidden layer along the vector difference between the true and alternative label gradients. This targeted approach ensures rigorous privacy guarantees while maintaining high model utility. 
Marvell~\cite{li2021label} mitigates label leakage by strategically perturbing cut-layer gradients, employing an optimized noise structure that minimizes the symmetric KL divergence between class-conditional gradient distributions.
Several research~\cite{wu2023federated, qiu2023evaluating} efforts secure the backward pass through the clipping of cut layer gradients and the subsequent injection of DP noise. By adding Laplace or Gaussian noise to these clipped gradients, these methods effectively thwart label inference attacks while theoretically maintaining the convergence of model training. 
To further refine this approach, ProjPert~\cite{fu2024projpert} formulates noise injection as a mathematical projection problem under privacy constraints, computing the optimal noise to minimize utility loss measured by Wasserstein distance.
Label DP~\cite{qiu2023evaluating, yang2022differentially} is implemented using a randomized response algorithm that randomly flips labels during training based on a specific probability, thereby preventing an attacker from perfectly reconstructing users' private labels.
Building upon this, PSLF introduces a Flipped Multi-Label Generation mechanism and two separate top sub-models, training one on flipped multi-labels to thwart label leakage and the other on true binary labels to boost prediction accuracy.

\subsubsection{Learning-based defenses} 
Learning-based defenses integrate privacy constraints directly into the optimization process via auxiliary training objectives or specialized loss functions. Unlike perturbation-based approaches that treat noise injection as a post-processing step, these strategies optimize internal representations or gradients to proactively minimize the leakage of sensitive information. 
NoPeek$^*$~\cite{vepakomma2020nopeek, shi2025navigating} reduces the risk of sensitive data leakage by incorporating distance correlation as a regularization term in the loss function to minimize the statistical dependency and achieve decorrelation between raw input and smashed data.
Building upon the core objective of dependency minimization, Zou et al.~\cite{zou2023mutual} introduced a Mutual Information Regularization~\cite{wang2021improving} term during training to effectively resist data reconstruction attacks and backdoor attacks by limiting the mutual information between private raw data and intermediate outputs while maintaining model utility.
\llm{Eguard$^*$~\cite{liu2024mitigating}} utilizes a projection network to migrate original embeddings into a secured space. It decouples the sensitive correlation between raw text and vectors through mutual information optimization, while a functionality preservation module ensures downstream task performance remains highly consistent with the original embeddings.
\llm{TextObfuscator$^*$~\cite{zhou2023textobfuscator}} identifies task-related prototypes through clustering and utilizes clustering and distance losses to group functionally similar representations while maintaining inter-prototype separation. By applying random perturbations within these clusters, it renders word vectors indistinguishable to protect privacy while preserving their original task functionality.
ResSFL~\cite{li2022ressfl} mitigates model inversion attacks by pre-training a resistant feature extractor via attacker-aware training and bottleneck layers, which is then transferred to initialize client models, maintaining high accuracy while reducing computational overhead and avoiding early-epoch vulnerabilities.
\llm{DualGuard$^*$~\cite{liu2025dualguard}} introduces a bidirectional defense for split-based LLM fine-tuning that uses local warm-up and global retention to transform the parameter space. By increasing the dissimilarity between client-side and pre-trained models, it successfully defends against both forward activation-based and backward gradient-based reconstruction attacks while preserving model utility.

\subsubsection{Detection-based defenses} 
Detection-based defenses focus on monitoring the training process to identify and respond to malicious activities in real-time. 

\noindent \textbf{Gradient attribute monitoring.}
This category of defense aims to prevent a malicious server from providing forged gradients to hijack the client's feature space to reconstruct original inputs.
A representative approach is SplitGuard~\cite{erdogan2022splitguard}, which detects training-hijacking attacks by leveraging the distinctively larger gradient angles produced when clients intermittently process "fake batches" with randomized labels. By analyzing these gradient discrepancies, SplitGuard enables clients to identify hijacked models and halt training before significant privacy leakage occurs. However, SplitGuard is vulnerable to the SplitSpy attack~\cite{fu2023focusing}, where a malicious server leverages a shadow model to identify fake samples and generate deceptive gradients. To address this, Fu et al. proposed Gradients Scrutinizer~\cite{fu2023focusing}, a passive monitoring mechanism that distinguishes honest training from hijacking by analyzing intrinsic gradient attributes: set gap, fitting Error, and overlapping ratio. This approach is robust against evasion and preserves model accuracy by avoiding the need for active sample injection.
SplitOut~\cite{erdougan2024splitout} uses the Local Outlier Factor (LOF) algorithm to monitor server-returned gradients in real-time, identifying malicious training-hijacking as outliers to intercept and terminate the attack at an early stage.

\noindent \textbf{Model parameter monitoring.} 
This strategy is designed to identify and prevent malicious clients from conducting backdoor attacks, where attackers attempt to implant hidden triggers into the model's parameters during the collaborative training process.
SafeSplit~\cite{rieger2025safesplit} determines whether a client has initiated an attack by monitoring the parameter updates of the server-side backbone before and after each client's training. Specifically, it utilizes the Discrete Cosine Transform (DCT) to monitor abnormal fluctuations of model parameters in the frequency domain as static analysis. For dynamic analysis, it introduces a rotational distance metric to capture directional mutations and angular velocity anomalies of the update trajectory in multi-dimensional space. If a client is discovered to be non-benign, the server determines that the training process is non-smooth and malicious. Subsequently, it activates a rollback mechanism.

\subsubsection{Split Unlearning} 
Machine unlearning~\cite{bourtoule2021machine} is essential for removing private or harmful data from trained models. However, the deep client-server coupling in traditional SL hinders efficient unlearning. To address this, SPLITWIPER~\cite{jiang2025splitunlearning} introduces a novel "one-way-one-off" propagation scheme. Clients freeze their weights after lightweight local pre-training and transmit intermediate outputs to the server only once. This decouples the network, satisfying the "Isolated" requirement of the SISA unlearning framework. Consequently, unlearning requests only require the target client to retrain, leaving others unaffected and enabling independent unlearning in multi-client SL architectures.

\subsubsection{Other Defenses} 

\noindent \textbf{Homomorphic encryption.} 
Homomorphic Encryption (HE) ensures zero privacy leakage by executing computations entirely on ciphertexts. Although HE has been actively explored in standard Split Learning SL~\cite{ pereteanu2022splithe,kanpak2024cure,liang2025dahe} and LLM inference~\cite{chen2022thex, lu2023bumblebee, hou2026ciphergpt}, its prohibitive computational and communication overheads currently preclude its practical application in split-based LLM fine-tuning.

\noindent \textbf{Protocol modifications.} 
Several studies preserve privacy by modifying the SL protocol.
3-Tier SL~\cite{alromih2022privacy} introduces a "Three-Tier Split Learning" architecture, adding an "Aggregator" as an intermediary between clients and the central server. Clients send only their activations to the aggregator, which averages them before forwarding them to the server, successfully preventing direct raw data leakage. However, this inevitably compromises model performance.
P-SL~\cite{pham2025split} is designed to prevent colluding malicious clients and servers from using shared local weights to reconstruct victims' private data via model inversion attacks. Consequently, P-SL fundamentally prohibits local weight sharing among clients, resulting in a unique local model for each client (similar to parallel split learning). Yet, because its primary focus is preventing client-server collusion, it overlooks the risk of a malicious server directly executing smashed data inversion attacks using methods like those mentioned above (Section~\autoref{subsubsection:smdia}).
\section{Conclusion}
\label{section: conclusion}
This survey provides a comprehensive review of recent advancements in integrating Split Learning with LLM Fine-Tuning (Split-based LLM-FT) , systematically exploring model-level optimizations, system-level performance enhancements, and privacy-preserving mechanisms. These components are abstracted into a cohesive, highly granular unified training pipeline : system-level optimizations, such as communication reduction and server scheduling, resolve critical bottlenecks and resource imbalances to ensure scalable execution ; model-level strategies tackle data heterogeneity, server-client update imbalances, and catastrophic forgetting to guarantee robust convergence and global accuracy ; and advanced privacy and security defenses safeguard sensitive intermediate representations against a diverse landscape of inversion, inference, and manipulation attacks. By systematically addressing these fundamental challenges across efficiency, performance, and security , this survey offers valuable insights for researchers and practitioners , paving the way for accessible, resource-efficient, and secure collaborative AI solutions that unlock the full potential of customized LLMs.



\bibliographystyle{ACM-Reference-Format}
\bibliography{sample-base}

@String{Computing = "Computing" }

@String{Computer = "{IEEE} Computer" }

@String{Springer = "Springer-Verlag" }

@ArtifactSoftware{R,
    title = {R: A Language and Environment for Statistical Computing},
    author = {{R Core Team}},
    organization = {R Foundation for Statistical Computing},
    address = {Vienna, Austria},
    year = {2019},
    url = {https://www.R-project.org/},
}

@inproceedings{he2019model,
  title={Model inversion attacks against collaborative inference},
  author={He, Zecheng and Zhang, Tianwei and Lee, Ruby B},
  booktitle={Proceedings of the 35th annual computer security applications conference},
  pages={148--162},
  year={2019},
  organization={ACM},
  publisher = {Association for Computing Machinery},
  address = {New York, NY, USA},
  series = {ACSAC '19}
}

@misc{cao2024sfprompt,
  title={Sfprompt: Communication-efficient split federated fine-tuning for large pre-trained models over resource-limited devices},
  author={Cao, Linxiao and Zhu, Yifei and Gong, Wei},
  journal={arXiv preprint arXiv:2407.17533},
  year={2024},
  eprint={2407.17533},
  archivePrefix={arXiv},
  url={https://arxiv.org/abs/2407.17533}, 
}

@inproceedings{lyu2023optimal,
  title={Optimal resource allocation for u-shaped parallel split learning},
  author={Lyu, Song and Lin, Zheng and Qu, Guanqiao and Chen, Xianhao and Huang, Xiaoxia and Li, Pan},
  booktitle={2023 IEEE Globecom Workshops (GC Wkshps)},
  pages={197--202},
  year={2023},
  organization={IEEE}
}

@article{wu2023split,
  title={Split learning over wireless networks: Parallel design and resource management},
  author={Wu, Wen and Li, Mushu and Qu, Kaige and Zhou, Conghao and Shen, Xuemin and Zhuang, Weihua and Li, Xu and Shi, Weisen},
  journal={IEEE Journal on Selected Areas in Communications},
  volume={41},
  number={4},
  pages={1051--1066},
  year={2023},
  publisher={IEEE}
}

@inproceedings{zhao2025efficient,
  title={SflLLM: Efficient Split Federated Learning for Large Language Model over Wireless Networks},
  author={Zhao, Kai and Zhu, Chen and Chen, Mingzhe and Huang, Chongwen and Yang, Zhaohui and Zhang, Zhaoyang},
  booktitle={GLOBECOM 2025-2025 IEEE Global Communications Conference},
  pages={1835--1840},
  year={2025},
  organization={IEEE}
}

@article{liang2025communication,
  author={Liang, Yipeng and Chen, Qimei and Li, Rongpeng and Zhu, Guangxu and Kaleem Awan, Muhammad and Jiang, Hao},
  journal={IEEE Transactions on Wireless Communications}, 
  title={Communication-and-Computation Efficient Split Federated Learning in Wireless Networks: Gradient Aggregation and Resource Management}, 
  year={2026},
  volume={25},
  number={},
  pages={1981-1995},
}

@article{lin2025hasfl,
  title={HASFL: Heterogeneity-aware split federated learning over edge computing systems},
  author={Lin, Zheng and Chen, Zhe and Chen, Xianhao and Ni, Wei and Gao, Yue},
  journal={IEEE Transactions on Mobile Computing},
  year={2026},
  publisher={IEEE}
}

@article{xu2023accelerating,
  title={Accelerating split federated learning over wireless communication networks},
  author={Xu, Ce and Li, Jinxuan and Liu, Yuan and Ling, Yushi and Wen, Miaowen},
  journal={IEEE Transactions on Wireless Communications},
  volume={23},
  number={6},
  pages={5587--5599},
  year={2023},
  publisher={IEEE}
}

@inproceedings{zhang2024resource,
  title={Resource-efficient Parallel Split Learning in Heterogeneous Edge Computing},
  author={Zhang, Mingjin and Cao, Jiannong and Sahni, Yuvraj and Chen, Xiangchun and Jiang, Shan},
  booktitle={2024 International Conference on Computing, Networking and Communications (ICNC)},
  pages={794--798},
  year={2024},
  organization={IEEE}
}

@article{oh2025communication,
  title={Communication-efficient split learning via adaptive feature-wise compression},
  author={Oh, Yongjeong and Lee, Jaeho and Brinton, Christopher G and Jeon, Yo-Seb},
  journal={IEEE Transactions on Neural Networks and Learning Systems},
  year={2025},
  publisher={IEEE}
}

@misc{wang2022fedlite,
  title={FedLite: A Scalable Approach for Federated Learning on Resource-constrained Clients}, 
  author={Jianyu Wang and Hang Qi and Ankit Singh Rawat and Sashank Reddi and Sagar Waghmare and Felix X. Yu and Gauri Joshi},
  year={2022},
  eprint={2201.11865},
  archivePrefix={arXiv},
  url={https://arxiv.org/abs/2201.11865}, 
}

@inproceedings{chen2021communication,
  title={Communication and computation reduction for split learning using asynchronous training},
  author={Chen, Xing and Li, Jingtao and Chakrabarti, Chaitali},
  booktitle={2021 IEEE Workshop on Signal Processing Systems (SiPS)},
  pages={76--81},
  year={2021},
  organization={IEEE},
}

@misc{shi2022efficient,
  title={An Efficient Split Fine-tuning Framework for Edge and Cloud Collaborative Learning}, 
  author={Shaohuai Shi and Qing Yang and Yang Xiang and Shuhan Qi and Xuan Wang},
  year={2022},
  eprint={2211.16703},
  archivePrefix={arXiv},
  url={https://arxiv.org/abs/2211.16703}, 
}

@inproceedings{hsieh2022c3,
  title={C3-SL: Circular convolution-based batch-wise compression for communication-efficient split learning},
  author={Hsieh, Cheng-Yen and Chuang, Yu-Chuan and Wu, An-Yeu},
  booktitle={IEEE 32nd International Workshop on Machine Learning for Signal Processing},
  pages={1--6},
  year={2022},
  organization={IEEE},
}

@inproceedings{shu2024dynsplit,
  title={Dynsplit: A dynamic split learning scheme for 5g-enpowered metaverse},
  author={Shu, Yunmeng and Gu, Pengwenlong and Adjih, C{\'e}dric and Chen, Chung Shue and Serhrouchni, Ahmed},
  booktitle={2024 IEEE International Conference on Metaverse Computing, Networking, and Applications (MetaCom)},
  pages={214--221},
  year={2024},
  organization={IEEE}
}

@misc{pal2021server,
  title={Server-Side Local Gradient Averaging and Learning Rate Acceleration for Scalable Split Learning}, 
  author={Shraman Pal and Mansi Uniyal and Jihong Park and Praneeth Vepakomma and Ramesh Raskar and Mehdi Bennis and Moongu Jeon and Jinho Choi},
  year={2021},
  eprint={2112.05929},
  archivePrefix={arXiv},
  url={https://arxiv.org/abs/2112.05929}, 
}

@article{lin2024split,
  title={Split learning in 6G edge networks},
  author={Lin, Zheng and Qu, Guanqiao and Chen, Xianhao and Huang, Kaibin},
  journal={IEEE Wireless Communications},
  volume={31},
  number={4},
  pages={170--176},
  year={2024},
  publisher={IEEE}
}

@inproceedings{tirana2024workflow,
  title={Workflow optimization for parallel split learning},
  author={Tirana, Joana and Tsigkari, Dimitra and Iosifidis, George and Chatzopoulos, Dimitris},
  booktitle={IEEE INFOCOM 2024-IEEE Conference on Computer Communications},
  pages={1331--1340},
  year={2024},
  organization={IEEE}
}

@misc{lin2024splitlora,
  title={SplitLoRA: A Split Parameter-Efficient Fine-Tuning Framework for Large Language Models}, 
  author={Zheng Lin and Xuanjie Hu and Yuxin Zhang and Zhe Chen and Zihan Fang and Xianhao Chen and Ang Li and Praneeth Vepakomma and Yue Gao},
  year={2024},
  eprint={2407.00952},
  archivePrefix={arXiv},
  url={https://arxiv.org/abs/2407.00952}, 
}

@misc{lin2025hsplitlora,
  title={HSplitLoRA: A Heterogeneous Split Parameter-Efficient Fine-Tuning Framework for Large Language Models}, 
  author={Zheng Lin and Yuxin Zhang and Zhe Chen and Zihan Fang and Xianhao Chen and Praneeth Vepakomma and Wei Ni and Jun Luo and Yue Gao},
  year={2025},
  eprint={2505.02795},
  archivePrefix={arXiv},
  url={https://arxiv.org/abs/2505.02795}, 
}

@inproceedings{li2024adaptive,
  title={Adaptive split learning over energy-constrained wireless edge networks},
  author={Li, Zuguang and Wu, Wen and Wu, Shaohua and Wang, Wei},
  booktitle={IEEE Conference on Computer Communications Workshops (INFOCOM WKSHPS)},
  pages={1--6},
  year={2024},
  organization={IEEE},
}

@article{oh2023mix2sfl,
  title={Mix2SFL: Two-way mixup for scalable, accurate, and communication-efficient split federated learning},
  author={Oh, Seungeun and Nam, Hyelin and Park, Jihong and Vepakomma, Praneeth and Raskar, Ramesh and Bennis, Mehdi and Kim, Seong-Lyun},
  journal={IEEE Transactions on Big Data},
  volume={10},
  number={3},
  pages={238--248},
  year={2023},
  publisher={IEEE}
}

@article{ao2025semi,
  author={Ao, Huiqing and Tian, Hui and Ni, Wanli and Nie, Gaofeng and Niyato, Dusit},
  journal={IEEE Transactions on Wireless Communications}, 
  title={Semi-Asynchronous Federated Split Learning for Computing-Limited Devices in Wireless Networks}, 
  year={2025},
  volume={24},
  number={6},
  pages={5196-5212},
}

@article{zhang2024pipar,
  title={PiPar: Pipeline parallelism for collaborative machine learning},
  author={Zhang, Zihan and Rodgers, Philip and Kilpatrick, Peter and Spence, Ivor and Varghese, Blesson},
  journal={Journal of Parallel and Distributed Computing},
  volume={193},
  pages={104947},
  year={2024},
  publisher={Elsevier}
}

@article{stephanie2023digital,
  title={Digital twin enabled asynchronous SplitFed learning in E-healthcare systems},
  author={Stephanie, Veronika and Khalil, Ibrahim and Atiquzzaman, Mohammed},
  journal={IEEE Journal on Selected Areas in Communications},
  volume={41},
  number={11},
  pages={3650--3661},
  year={2023},
  publisher={IEEE}
}

@inproceedings{chopra2021adasplit,
  title={Adaptive split learning},
  author={Chopra, Ayush and Sahu, Surya Kant and Singh, Abhishek and Java, Abhinav and Vepakomma, Praneeth and Amiri, Mohammad Mohammadi and Raskar, Ramesh},
  booktitle={Federated Learning Systems (FLSys) Workshop@ MLSys 2023},
  year={2023},
}

@misc{wu2025sfl,
  title={SFL-LEO: Asynchronous Split-Federated Learning Design for LEO Satellite-Ground Network Framework}, 
  author={Jiasheng Wu and Jingjing Zhang and Zheng Lin and Zhe Chen and Xiong Wang and Wenjun Zhu and Yue Gao},
  year={2025},
  eprint={2504.13479},
  archivePrefix={arXiv},
  url={https://arxiv.org/abs/2504.13479}, 
}

@inproceedings{erdougan2022unsplit,
  title={Unsplit: Data-oblivious model inversion, model stealing, and label inference attacks against split learning},
  author={Erdo{\u{g}}an, Ege and K{\"u}p{\c{c}}{\"u}, Alptekin and {\c{C}}i{\c{c}}ek, A Erc{\"u}ment},
  booktitle={Proceedings of the 21st Workshop on Privacy in the Electronic Society},
  pages={115--124},
  year={2022},
  publisher = {Association for Computing Machinery},
  address = {New York, NY, USA},
  numpages = {10},
}

@inproceedings{han2023splitgp,
  title={SplitGP: Achieving both generalization and personalization in federated learning},
  author={Han, Dong-Jun and Kim, Do-Yeon and Choi, Minseok and Brinton, Christopher G and Moon, Jaekyun},
  booktitle={IEEE Conference on Computer Communications},
  pages={1--10},
  year={2023},
  organization={IEEE},
}

@inproceedings{zhu2024dualfed,
  title={Dualfed: enjoying both generalization and personalization in federated learning via hierachical representations},
  author={Zhu, Guogang and Liu, Xuefeng and Niu, Jianwei and Tang, Shaojie and Wu, Xinghao and Zhang, Jiayuan},
  booktitle={Proceedings of the 32nd ACM International Conference on Multimedia},
  pages={11060--11069},
  year={2024},
  address = {New York, NY, USA},
}

@article{han2023federated,
  title={Federated split learning with joint personalization-generalization for inference-stage optimization in wireless edge networks},
  author={Han, Dong-Jun and Kim, Do-Yeon and Choi, Minseok and Nickel, David and Moon, Jaekyun and Chiang, Mung and Brinton, Christopher G},
  journal={IEEE Transactions on Mobile Computing},
  volume={23},
  number={6},
  pages={7048--7065},
  year={2023},
  publisher={IEEE},
}

@inproceedings{liao2024parallelsfl,
  title={Parallelsfl: A novel split federated learning framework tackling heterogeneity issues},
  author={Liao, Yunming and Xu, Yang and Xu, Hongli and Yao, Zhiwei and Huang, Liusheng and Qiao, Chunming},
  booktitle={Proceedings of the 30th Annual International Conference on Mobile Computing and Networking},
  pages={845--860},
  year={2024},
  address = {New York, NY, USA},
  pages = {845–860},
  numpages = {16},
}

@inproceedings{yang2025gas,
  title={GAS: Generative Activation-Aided Asynchronous Split Federated Learning},
  author={Yang, Jiarong and Liu, Yuan},
  booktitle={Proceedings of the AAAI Conference on Artificial Intelligence},
  volume={39},
  number={20},
  pages={21956--21964},
  year={2025}
}

@misc{yang2024scala,
  title={SCALA: Split Federated Learning with Concatenated Activations and Logit Adjustments}, 
  author={Jiarong Yang and Yuan Liu},
  year={2025},
  eprint={2405.04875},
  archivePrefix={arXiv},
  url={https://arxiv.org/abs/2405.04875}, 
}

@article{yao2025pairingfl,
  title={PairingFL: Efficient Federated Learning With Model Splitting and Client Pairing},
  author={Yao, Zhiwei and Qi, Ji and Xu, Yang and Liao, Yunming and Xu, Hongli and Wang, Lun},
  journal={IEEE Transactions on Networking},
  year={2025},
  publisher={IEEE}
}

@article{chen2023personalized,
  title={Personalized fair split learning for resource-constrained Internet of Things},
  author={Chen, Haitian and Chen, Xuebin and Peng, Lulu and Bai, Yuntian},
  journal={Sensors},
  volume={24},
  number={1},
  pages={88},
  year={2023},
  publisher={MDPI},
}

@article{tinh2025smixsl,
  title={SMixSL: The Smashed-Mixture Technique for Split Learning With Localizable Features},
  author={Tinh, Vo Phuc and Khoa, Tran Anh and Lam, Pham Duc and Nam, Nguyen Hoang and Dang, Duc Ngoc Minh and Le, Duy-Dong and Dang, Thai-Thinh and Nguyen, Van-Luong and Pham, Thanh-Qui and Nguyen, Thai-Binh},
  journal={IEEE Transactions on Emerging Topics in Computational Intelligence},
  year={2025},
  publisher={IEEE}
}

@misc{yuan2025flexpsfl,
  title={Flexible Personalized Split Federated Learning for On-Device Fine-Tuning of Foundation Models}, 
  author={Tianjun Yuan and Jiaxiang Geng and Pengchao Han and Xianhao Chen and Bing Luo},
  year={2025},
  eprint={2508.10349},
  archivePrefix={arXiv},
  url={https://arxiv.org/abs/2508.10349}, 
}

@article{chawla2024beyond,
  title={Beyond Federated Learning for IoT: Efficient Split Learning With Caching and Model Customization},
  author={Chawla, Manisha and Gupta, Gagan Raj and Gaddam, Shreyas and Wadhwa, Manas},
  journal={IEEE Internet of Things Journal},
  volume={11},
  number={20},
  pages={32617--32630},
  year={2024},
  publisher={IEEE},
}

@inproceedings{xia2025multisfl,
  title={MultiSFL: Towards Accurate Split Federated Learning via Multi-Model Aggregation and Knowledge Replay},
  author={Xia, Zeke and Hu, Ming and Yan, Dengke and Liu, Ruixuan and Li, Anran and Xie, Xiaofei and Chen, Mingsong},
  booktitle={Proceedings of the AAAI Conference on Artificial Intelligence},
  volume={39},
  number={1},
  pages={914--922},
  year={2025}
}

@article{hu2025review,
  title={A review and experimental evaluation on split learning},
  author={Hu, Zhanyi and Zhou, Tianchen and Wu, Bingzhe and Chen, Cen and Wang, Yanhao},
  journal={Future Internet},
  volume={17},
  number={2},
  pages={87},
  year={2025},
  publisher={MDPI},
}

@inproceedings{khan2025oops,
  title={Oops!... They Stole it Again: Attacks on Split Learning},
  author={Khan, Tanveer and Michalas, Antonis},
  booktitle={Proceedings of the 18th ACM Workshop on Artificial Intelligence and Security},
  pages={123--135},
  year={2026},
  numpages = {13},
  publisher = {Association for Computing Machinery},
  address = {New York, NY, USA},
}

@misc{shabbir2025taxonomy,
  title={A Taxonomy of Attacks and Defenses in Split Learning}, 
  author={Aqsa Shabbir and Halil İbrahim Kanpak and Alptekin Küpçü and Sinem Sav},
  year={2025},
  eprint={2505.05872},
  archivePrefix={arXiv},
  url={https://arxiv.org/abs/2505.05872}, 
}

@inproceedings{hu2025slperf,
  title={SLPerf: A Research Library and Benchmark Framework for Split Learning},
  author={Hu, Zhanyi and Zhou, Tianchen and Wu, Bingzhe and Chen, Cen and Wang, Yanhao},
  booktitle={2025 IEEE 41st International Conference on Data Engineering Workshops (ICDEW)},
  pages={33--36},
  year={2025},
  organization={IEEE},
}

@inproceedings{radovivc2025towards,
  title={Towards a Unified Framework for Split Learning},
  author={Radovi{\v{c}}, Boris and Canini, Marco and Horv{\'a}th, Samuel and Pejovi{\'c}, Veljko and Vepakomma, Praneeth},
  booktitle={Proceedings of the 5th Workshop on Machine Learning and Systems},
  pages={183--191},
  year={2025}
}

@inproceedings{gu2025vflair,
  title={VFLAIR-LLM: A Comprehensive Framework and Benchmark for Split Learning of LLMs},
  author={Gu, Zixuan and Fan, Qiufeng and Sun, Long and Liu, Yang and Ye, Xiaojun},
  booktitle={Proceedings of the 31st ACM SIGKDD Conference on Knowledge Discovery and Data Mining V. 2},
  pages={5470--5481},
  year={2025},
  address = {New York, NY, USA},
  series = {KDD '25},
  publisher = {Association for Computing Machinery},
}

@article{duan2022combined,
  title={Combined federated and split learning in edge computing for ubiquitous intelligence in internet of things: State-of-the-art and future directions},
  author={Duan, Qiang and Hu, Shijing and Deng, Ruijun and Lu, Zhihui},
  journal={Sensors},
  volume={22},
  number={16},
  pages={5983},
  year={2022},
  publisher={MDPI},
}

@article{hukkeri2025comprehensive,
  title={A Comprehensive Survey on Split-Fed Learning: Methods, Innovations, and Future Directions},
  author={Hukkeri, Geetabai S and Goudar, RH and Dhananjaya, GM and Rathod, Vijayalaxmi N and Ankalaki, Shilpa},
  journal={IEEE Access},
  volume={13},
  pages={46312--46333},
  year={2025},
}

@article{gupta2018distributed,
  title = {Distributed learning of deep neural network over multiple agents},
  journal = {Journal of Network and Computer Applications},
  volume = {116},
  pages = {1-8},
  year = {2018},
  issn = {1084-8045},
  author = {Otkrist Gupta and Ramesh Raskar},
}

@misc{vepakomma2018split,
  title={Split learning for health: Distributed deep learning without sharing raw patient data}, 
  author={Praneeth Vepakomma and Otkrist Gupta and Tristan Swedish and Ramesh Raskar},
  year={2018},
  eprint={1812.00564},
  archivePrefix={arXiv},
  url={https://arxiv.org/abs/1812.00564}, 
}

@inproceedings{jeon2020privacy,
  title={Privacy-sensitive parallel split learning},
  author={Jeon, Joohyung and Kim, Joongheon},
  booktitle={2020 International Conference on Information Networking (ICOIN)},
  pages={7--9},
  year={2020},
  organization={IEEE},
}

@article{gao2024pipesfl,
  title={PipeSFL: A fine-grained parallelization framework for split federated learning on heterogeneous clients},
  author={Gao, Yunqi and Hu, Bing and Mashhadi, Mahdi Boloursaz and Wang, Wei and Bennis, Mehdi},
  journal={IEEE transactions on mobile computing},
  publisher={IEEE},
  year={2025},
  volume={24},
  number={3},
  pages={1774-1791},
}

@inproceedings{chen2024unveiling,
  title={Unveiling the vulnerability of private fine-tuning in split-based frameworks for large language models: A bidirectionally enhanced attack},
  author={Chen, Guanzhong and Qin, Zhenghan and Yang, Mingxin and Zhou, Yajie and Fan, Tao and Du, Tianyu and Xu, Zenglin},
  booktitle={Proceedings of the 2024 on ACM SIGSAC Conference on Computer and Communications Security},
  pages={2904--2918},
  year={2024},
  address = {New York, NY, USA},
}

@article{shen2023ringsfl,
  title={Ringsfl: An adaptive split federated learning towards taming client heterogeneity},
  author={Shen, Jinglong and Cheng, Nan and Wang, Xiucheng and Lyu, Feng and Xu, Wenchao and Liu, Zhi and Aldubaikhy, Khalid and Shen, Xuemin},
  journal={IEEE Transactions on Mobile Computing},
  volume={23},
  number={5},
  pages={5462--5478},
  year={2023},
  publisher={IEEE}
}

@inproceedings{zhang2024splitllm,
  title={Splitllm: Hierarchical split learning for large language model over wireless network},
  author={Zhang, Songge and Cheng, Guoliang and Li, Zuguang and Wu, Wen},
  booktitle={2024 IEEE Globecom Workshops (GC Wkshps)},
  pages={1--6},
  year={2024},
  organization={IEEE}
}

@article{lin2025hierarchical,
  title={Hierarchical split federated learning: Convergence analysis and system optimization},
  author={Lin, Zheng and Wei, Wei and Chen, Zhe and Lam, Chan-Tong and Chen, Xianhao and Gao, Yue and Luo, Jun},
  journal={IEEE Transactions on Mobile Computing},
  year={2025},
  publisher={IEEE}
}

@article{fan2025madrl,
  title={MADRL-based model partitioning, aggregation control, and resource allocation for cloud-edge-device collaborative split federated learning},
  author={Fan, Wenhao and Chen, Penghui and Chun, Xiongfei and Liu, Yuan'an},
  journal={IEEE Transactions on Mobile Computing},
  year={2025},
  publisher={IEEE},
}

@article{lin2024efficient,
  title={Efficient parallel split learning over resource-constrained wireless edge networks},
  author={Lin, Zheng and Zhu, Guangyu and Deng, Yiqin and Chen, Xianhao and Gao, Yue and Huang, Kaibin and Fang, Yuguang},
  journal={IEEE Transactions on Mobile Computing},
  volume={23},
  number={10},
  pages={9224--9239},
  year={2024},
  publisher={IEEE}
}

@article{samikwa2022ares,
  title={Ares: Adaptive resource-aware split learning for internet of things},
  author={Samikwa, Eric and Di Maio, Antonio and Braun, Torsten},
  journal={Computer Networks},
  volume={218},
  pages={109380},
  year={2022},
  publisher={Elsevier}
}

@article{marinova2025optimal,
  title={Optimal Cut Layer Bounds for Split Learning},
  author={Marinova, Matea and Poposka, Marija and Hadzi-Velkov, Zoran and Rakovic, Valentin},
  journal={IEEE Communications Letters},
  year={2025},
  publisher={IEEE}
}

@inproceedings{trung2025latency,
  title={Latency-Aware Split Learning Optimization via Genetic Algorithms},
  author={Trung, Le Hoang and Nguyen, Tan Y and Le, Duy Dong and Dang, Thai Thinh and Khoa, Tran Anh},
  booktitle={Proceedings of the 6th Workshop on Intelligent Cross-Data Analysis and Retrieval},
  pages={14--19},
  year={2025}
}

@inproceedings{chen2025memory,
  title={Memory-efficient split federated learning for llm fine-tuning on heterogeneous mobile devices},
  author={Chen, Xiaopei and Li, Liang and Ji, Fei and Wu, Wen},
  booktitle={IEEE INFOCOM 2025-IEEE Conference on Computer Communications Workshops (INFOCOM WKSHPS)},
  pages={1--6},
  year={2025},
  organization={IEEE},
}

@article{ma2025splitfrozen,
  title={SplitFrozen: Split learning with device-side model frozen for fine-tuning LLM on heterogeneous resource-constrained devices},
  author={Ma, Jian and Lyu, Xinchen and Jiang, Jun and Cui, Qimei and Yao, Haipeng and Tao, Xiaofeng},
  journal={IEEE Communications Magazine},
  year={2025},
  publisher={IEEE}
}

@inproceedings{zheng2023reducing,
  title={Reducing communication for split learning by randomized top-k sparsification},
  author={Zheng, Fei and Chen, Chaochao and Lyu, Lingjuan and Yao, Binhui},
  booktitle={Proceedings of the Thirty-Second International Joint Conference on Artificial Intelligence},
  pages={4665--4673},
  year={2023}
}

@inproceedings{zhou2024mask,
    title       = {Mask-encoded sparsification: Overcoming biased gradients for communication-efficient split learning},
    author      = {Wenxuan Zhou and Zhihao Qu and Shen-Huan Lyu and Miao Cai and Baoliu Ye},
    booktitle   = {Proceedings of the 27th European Conference on Artificial Intelligence},
    pages       = {2806--2813},
    year        = {2024}
}

@inproceedings{gao2025communication,
  title={Communication-Efficient Split Federated Learning with Dynamic Feature Compression},
  author={Gao, Siyu},
  booktitle={2025 IEEE 6th International Seminar on Artificial Intelligence, Networking and Information Technology (AINIT)},
  pages={01--07},
  year={2025},
  organization={IEEE},
}

@inproceedings{shao2020bottlenet++,
  title={Bottlenet++: An end-to-end approach for feature compression in device-edge co-inference systems},
  author={Shao, Jiawei and Zhang, Jun},
  booktitle={2020 IEEE International Conference on Communications Workshops (ICC Workshops)},
  pages={1--6},
  year={2020},
  organization={IEEE}
}

@article{lin2025slacc,
  title={SL-ACC: A Communication-Efficient Split Learning Framework with Adaptive Channel-wise Compression},
  author={Zehang Lin and Zhe Lin and Miao Yang and Jianhao Huang and Yuxin Zhang and Zihan Fang and Xia Du and Zhe Chen and Shunzhi Zhu and Wei Ni},
  journal={IEEE Transactions on Vehicular Technology},
  year={2026},
}

@inproceedings{radovivc2024train,
  title={Train your cake and eat it too! Repurposing collaborative training to tailor LLMs to private data without sharing},
  author={Radovi{\v{c}}, Boris and Aljahdali, Mohammed and Canini, Marco and Pejovi{\'c}, Veljko and Khayyat, Zuhair},
  booktitle={Workshop on Efficient Systems for Foundation Models II@ ICML2024},
  year={2024}
}

@inproceedings{mu2023CSEFSL,
  title={Communication and storage efficient federated split learning},
  author={Mu, Yujia and Shen, Cong},
  booktitle={ICC 2023-IEEE International Conference on Communications},
  pages={2976--2981},
  year={2023},
  organization={IEEE}
}

@inproceedings{nairfsl,
  title={FSL-SAGE: Accelerating Federated Split Learning via Smashed Activation Gradient Estimation},
  author={Nair, Srijith and Lin, Michael and Ju, Peizhong and Talebi, Amirreza and Bentley, Elizabeth Serena and Liu, Jia},
  booktitle={Forty-second International Conference on Machine Learning},
  year={2025},
  location = {Vancouver, Canada},
}

@inproceedings{han2021accelerating,
  title={Accelerating federated learning with split learning on locally generated losses},
  author={Han, Dong-Jun and Bhatti, Hasnain Irshad and Lee, Jungmoon and Moon, Jaekyun},
  booktitle={ICML 2021 workshop on federated learning for user privacy and data confidentiality. ICML Board},
  year={2021}
}

@misc{zhang2025ampere,
  title={Ampere: Communication-Efficient and High-Accuracy Split Federated Learning}, 
  author={Zihan Zhang and Leon Wong and Blesson Varghese},
  year={2025},
  eprint={2507.07130},
  archivePrefix={arXiv},
  url={https://arxiv.org/abs/2507.07130}, 
}

@inproceedings{lin2025eslleo,
  title={Esl-leo: An efficient split learning framework over leo satellite networks},
  author={Lin, Zheng and Zhang, Yuxin and Chen, Zhe and Fang, Zihan and Yang, Yanni and Zhang, Guoming and Yang, Huan and Wu, Cong and Chen, Xianhao and Gao, Yue},
  booktitle={International Conference on Wireless Artificial Intelligent Computing Systems and Applications},
  pages={344--357},
  year={2025},
  organization={Springer}
}

@inproceedings{hu2024menos,
  title={Menos: Split Fine-Tuning Large Language Models with Efficient GPU Memory Sharing},
  author={Hu, Chenghao and Li, Baochun},
  booktitle={Proceedings of the 25th International Middleware Conference},
  pages={185--198},
  year={2024},
  publisher = {Association for Computing Machinery},
  address = {New York, NY, USA},
  numpages = {14},
}

@inproceedings{he2025hourglass,
  title={Hourglass: Enabling Efficient Split Federated Learning with Data Parallelism},
  author={He, Qiang and Wang, Kaibin and Dong, Zeqian and Yuan, Liang and Chen, Feifei and Jin, Hai and Yang, Yun},
  booktitle={Proceedings of the Twentieth European Conference on Computer Systems},
  pages={1317--1333},
  year={2025},
  publisher = {Association for Computing Machinery},
  address = {New York, NY, USA},
  series = {EuroSys '25},
  numpages = {17}
}

@misc{joshi2021splitfed,
  title={Splitfed learning without client-side synchronization: Analyzing client-side split network portion size to overall performance}, 
  author={Praveen Joshi and Chandra Thapa and Seyit Camtepe and Mohammed Hasanuzzamana and Ted Scully and Haithem Afli},
  year={2021},
  eprint={2109.09246},
  archivePrefix={arXiv},
  url={https://arxiv.org/abs/2109.09246}, 
}

@misc{liu2025communication,
  title={Communication-Computation Pipeline Parallel Split Learning over Wireless Edge Networks}, 
  author={Chenyu Liu and Zhaoyang Zhang and Zirui Chen and Zhaohui Yang},
  year={2025},
  eprint={2511.23167},
  archivePrefix={arXiv},
  url={https://arxiv.org/abs/2511.23167}, 
}

@article{liao2023accelerating,
  title={Accelerating federated learning with data and model parallelism in edge computing},
  author={Liao, Yunming and Xu, Yang and Xu, Hongli and Yao, Zhiwei and Wang, Lun and Qiao, Chunming},
  journal={IEEE/ACM Transactions on Networking},
  volume={32},
  number={1},
  pages={904--918},
  year={2023},
  publisher={IEEE}
}

@inproceedings{liao2024mergesfl,
  title={MergeSFL: Split federated learning with feature merging and batch size regulation},
  author={Liao, Yunming and Xu, Yang and Xu, Hongli and Wang, Lun and Yao, Zhiwei and Qiao, Chunming},
  booktitle={2024 IEEE 40th International Conference on Data Engineering (ICDE)},
  pages={2054--2067},
  year={2024},
  organization={IEEE}
}

@article{li2025energy,
  title={Energy-efficient split learning for fine-tuning large language models in edge networks},
  author={Li, Zuguang and Wu, Shaohua and Li, Liang and Zhang, Songge},
  journal={IEEE Networking Letters},
  year={2025},
  publisher={IEEE},
  volume={7},
  number={3},
  pages={176-180},
}

@article{lin2025adaptsfl,
  title={Adaptsfl: Adaptive split federated learning in resource-constrained edge networks},
  author={Lin, Zheng and Qu, Guanqiao and Wei, Wei and Chen, Xianhao and Leung, Kin K},
  journal={IEEE Transactions on Networking},
  year={2025},
  publisher={IEEE}
}

@inproceedings{sun2025split,
  title={Split federated learning over heterogeneous edge devices: Algorithm and optimization},
  author={Sun, Yunrui and Hu, Gang and Teng, Yinglei and Cai, Dunbo},
  booktitle={2025 IEEE Wireless Communications and Networking Conference (WCNC)},
  pages={01--06},
  year={2025},
  organization={IEEE}
}

@article{zhang2025split,
  title={Split fine-tuning for large language models in wireless networks},
  author={Zhang, Songge and Cheng, Guoliang and Wu, Wen and Huang, Xinyu and Song, Lingyang and Shen, Xuemin},
  journal={IEEE Journal of Selected Topics in Signal Processing},
  year={2025},
  publisher={IEEE}
}

@inproceedings{borzunov2023petals,
    title = "Petals: Collaborative Inference and Fine-tuning of Large Models",
    author={Borzunov, Alexander and Baranchuk, Dmitry and Dettmers, Tim and Riabinin, Maksim and Belkada, Younes and Chumachenko, Artem and Samygin, Pavel and Raffel, Colin},
    booktitle = "Proceedings of the 61st Annual Meeting of the Association for Computational Linguistics (Volume 3: System Demonstrations)",
    month = jul,
    year = "2023",
    address = "Toronto, Canada",
    publisher = "Association for Computational Linguistics",
    pages = "558--568",
}

@inproceedings{wadhwa2023pfsl,
  title={Pfsl: Personalized \& fair split learning with data \& label privacy for thin clients},
  author={Wadhwa, Manas and Gupta, Gagan Raj and Sahu, Ashutosh and Saini, Rahul and Mittal, Vidhi},
  booktitle={2023 IEEE/ACM 23rd International Symposium on Cluster, Cloud and Internet Computing (CCGrid)},
  pages={377--390},
  year={2023},
  organization={IEEE}
}

@article{luo2024fsmkd,
  title={Federated split learning via mutual knowledge distillation},
  author={Luo, Linjun and Zhang, Xinglin},
  journal={IEEE Transactions on Network Science and Engineering},
  volume={11},
  number={3},
  pages={2729--2741},
  year={2024},
  publisher={IEEE}
}

@misc{huang2023minibatch,
  title={When minibatch sgd meets splitfed learning: Convergence analysis and performance evaluation},
  author={Huang, Chao and Tian, Geng and Tang, Ming},
  archivePrefix={arXiv},
  year={2023},
  eprint={2308.11953},
  url={https://arxiv.org/abs/2308.11953}
}

@misc{yan2023s2fl,
  title={Have Your Cake and Eat It Too: Toward Efficient and Accurate Split Federated Learning}, 
  author={Dengke Yan and Ming Hu and Zeke Xia and Yanxin Yang and Jun Xia and Xiaofei Xie and Mingsong Chen},
  year={2024},
  eprint={2311.13163},
  archivePrefix={arXiv},
  url={https://arxiv.org/abs/2311.13163}, 
}

@inproceedings{oh2022locfedmix,
  title={Locfedmix-sl: Localize, federate, and mix for improved scalability, convergence, and latency in split learning},
  author={Oh, Seungeun and Park, Jihong and Vepakomma, Praneeth and Baek, Sihun and Raskar, Ramesh and Bennis, Mehdi and Kim, Seong-Lyun},
  booktitle={Proceedings of the ACM Web Conference 2022},
  pages={3347--3357},
  year={2022},
  location = {Virtual Event, Lyon, France},
}

@article{feng2024slwf,
  title={SLwF: A split learning without forgetting framework for Internet of Things},
  author={Feng, Xingyu and Jia, Renqi and Luo, Chengwen and Leung, Victor CM and Xu, Weitao},
  journal={IEEE Internet of Things Journal},
  year={2024},
  volume={12},
  number={9},
  pages={12008-12020},
  publisher={IEEE},
}

@inproceedings{pasquini2021unleashing,
  title={Unleashing the tiger: Inference attacks on split learning},
  author={Pasquini, Dario and Ateniese, Giuseppe and Bernaschi, Massimo},
  booktitle={Proceedings of the 2021 ACM SIGSAC conference on computer and communications security},
  pages={2113--2129},
  year={2021},
  address = {New York, NY, USA},
}

@preprint{gawron2022feature,
  title={Feature space hijacking attacks against differentially private split learning},
  author={Gawron, Grzegorz and Stubbings, Philip},
  year={2022},
  eprint={2201.04018},
  archivePrefix={arXiv},
  url={https://arxiv.org/abs/2201.04018}, 
}

@article{yu2024sia,
  title={SIA: A sustainable inference attack framework in split learning},
  author={Yu, Fangchao and Wang, Lina and Zeng, Bo and Zhao, Kai and Wu, Tian and Pang, Zhi},
  journal={Neural Networks},
  volume={171},
  pages={396--409},
  year={2024},
  publisher={Elsevier}
}

@inproceedings{erdogan2022splitguard,
  title={Splitguard: Detecting and mitigating training-hijacking attacks in split learning},
  author={Erdogan, Ege and K{\"u}p{\c{c}}{\"u}, Alptekin and Cicek, A Ercument},
  booktitle={Proceedings of the 21st Workshop on Privacy in the Electronic Society},
  pages={125--137},
  year={2022},
  publisher = {Association for Computing Machinery},
  address = {New York, NY, USA},
}

@inproceedings{fu2023focusing,
  title={Focusing on Pinocchio's Nose: A Gradients Scrutinizer to Thwart Split-Learning Hijacking Attacks Using Intrinsic Attributes.},
  author={Fu, Jiayun and Ma, Xiaojing and Zhu, Bin B and Hu, Pingyi and Zhao, Ruixin and Jia, Yaru and Xu, Peng and Jin, Hai and Zhang, Dongmei},
  booktitle={30th Annual Network and Distributed System Security Symposium},
  address={San Diego, California, USA},
  year={2023},
}

@inproceedings{gao2023pcat,
  title={$\{$PCAT$\}$: Functionality and data stealing from split learning by $\{$Pseudo-Client$\}$ attack},
  author={Gao, Xinben and Zhang, Lan},
  booktitle={32nd USENIX Security Symposium (USENIX Security 23)},
  pages={5271--5288},
  year={2023},
  address = {Anaheim, CA},
  publisher = {USENIX Association},
}

@article{zhang2024functionality,
  title={Functionality and data stealing by pseudo-client attack and target defenses in split learning},
  author={Zhang, Lan and Gao, Xinben and Li, Yaliang and Liu, Yunhao},
  journal={IEEE Transactions on Dependable and Secure Computing},
  volume={22},
  number={1},
  pages={84--100},
  year={2024},
  publisher={IEEE}
}

@article{zeng2025gan,
  title={GAN-based data reconstruction attacks in split learning},
  author={Zeng, Bo and Luo, Sida and Yu, Fangchao and Yang, Geying and Zhao, Kai and Wang, Lina},
  journal={Neural Networks},
  volume={185},
  pages={107150},
  year={2025},
  publisher={Elsevier}
}

@inproceedings{luo2023fsa,
  title={Feature sniffer: A stealthy inference attacks framework on split learning},
  author={Luo, Sida and Yu, Fangchao and Wang, Lina and Zeng, Bo and Pang, Zhi and Zhao, Kai},
  booktitle={International conference on artificial neural networks},
  pages={66--77},
  year={2023},
  organization={Springer},
  address="Cham"
}

@article{li2023gan,
  title={GAN you see me? enhanced data reconstruction attacks against split inference},
  author={Li, Ziang and Yang, Mengda and Liu, Yaxin and Wang, Juan and Hu, Hongxin and Yi, Wenzhe and Xu, Xiaoyang},
  journal={Advances in neural information processing systems},
  volume={36},
  pages={54554--54566},
  year={2023},
}

@inproceedings{xu2024stealthy,
  title={A stealthy wrongdoer: Feature-oriented reconstruction attack against split learning},
  author={Xu, Xiaoyang and Yang, Mengda and Yi, Wenzhe and Li, Ziang and Wang, Juan and Hu, Hongxin and Zhuang, Yong and Liu, Yaxin},
  booktitle={Proceedings of the IEEE/CVF conference on computer vision and pattern recognition},
  pages={12130--12139},
  year={2024},
  address = {Seattle, WA, USA}
}

@inproceedings{zhu2025passive,
  author       = {Xiaochen Zhu and
                  Xinjian Luo and
                  Yuncheng Wu and
                  Yangfan Jiang and
                  Xiaokui Xiao and
                  Beng Chin Ooi},
  title        = {Passive Inference Attacks on Split Learning via Adversarial Regularization},
  booktitle    = {Network and Distributed System Security (NDSS) Symposium 2025},
  address      = {San Diego, CA},
  year         = {2025},
  month        = {Feburary},
}

@inproceedings{liu2025dualguard,
  title={DualGuard: A Parameter Space Transformation Approach for Bidirectional Defense in Split-Based LLM Fine-Tuning},
  author={Liu, Zihan and Wang, Yizhen and Wang, Rui and Wu, Sai},
  booktitle={Proceedings of the 63rd Annual Meeting of the Association for Computational Linguistics (Volume 1: Long Papers)},
  pages={17065--17080},
  year={2025},
  address = "Vienna, Austria",
}

@inproceedings{li2021label,
  title={Label Leakage and Protection in Two-party Split Learning},
  author={Li, Oscar and Sun, Jiankai and Yang, Xin and Gao, Weihao and Zhang, Hongyi and Xie, Junyuan and Smith, Virginia and Wang, Chong},
  booktitle={International Conference on Learning Representations},
  year={2022},
}

@inproceedings{liu2023distance,
  title={Distance-based online label inference attacks against split learning},
  author={Liu, Junlin and Lyu, Xinchen},
  booktitle={ICASSP 2023-2023 IEEE International Conference on Acoustics, Speech and Signal Processing (ICASSP)},
  pages={1--5},
  year={2023},
  organization={IEEE}
}

@article{liu2024similarity,
  title={Similarity-based label inference attack against training and inference of split learning},
  author={Liu, Junlin and Lyu, Xinchen and Cui, Qimei and Tao, Xiaofeng},
  journal={IEEE Transactions on Information Forensics and Security},
  volume={19},
  pages={2881--2895},
  year={2024},
  publisher={IEEE}
}

@article{zhu2019deep,
  title={Deep leakage from gradients},
  author={Zhu, Ligeng and Liu, Zhijian and Han, Song},
  journal={Advances in neural information processing systems},
  volume={32},
  year={2019},
  address = {Red Hook, NY, USA},
}

@inproceedings{kariyappa2023exploit,
  title={Exploit: Extracting private labels in split learning},
  author={Kariyappa, Sanjay and Qureshi, Moinuddin K},
  booktitle={2023 IEEE conference on secure and trustworthy machine learning (SaTML)},
  pages={165--175},
  year={2023},
  organization={IEEE},
}

@misc{xie2023label,
  title={Label Inference Attack against Split Learning under Regression Setting}, 
  author={Shangyu Xie and Xin Yang and Yuanshun Yao and Tianyi Liu and Taiqing Wang and Jiankai Sun},
  year={2023},
  eprint={2301.07284},
  archivePrefix={arXiv},
  url={https://arxiv.org/abs/2301.07284}, 
}

@article{zhao2024splitaum,
  title={SplitAUM: Auxiliary model-based label inference attack against split learning},
  author={Zhao, Kai and Chuo, Xiaowei and Yu, Fangchao and Zeng, Bo and Pang, Zhi and Wang, Lina},
  journal={IEEE Transactions on Network and Service Management},
  year={2024},
  publisher={IEEE}
}

@inproceedings{fu2022label,
  title={Label inference attacks against vertical federated learning},
  author={Fu, Chong and Zhang, Xuhong and Ji, Shouling and Chen, Jinyin and Wu, Jingzheng and Guo, Shanqing and Zhou, Jun and Liu, Alex X and Wang, Ting},
  booktitle={31st USENIX security symposium (USENIX Security 22)},
  pages={1397--1414},
  year={2022}
}

@misc{zhao2020idlg,
  title={idlg: Improved deep leakage from gradients},
  author={Zhao, Bo and Mopuri, Konda Reddy and Bilen, Hakan},
  eprint={2301.07284},
  archivePrefix={arXiv},
  year={2020},
  url={http://arxiv.org/abs/2001.02610},
}

@inproceedings{balunovic2022lamp,
 author = {Balunovic, Mislav and Dimitrov, Dimitar and Jovanovi\'{c}, Nikola and Vechev, Martin},
 booktitle = {Advances in Neural Information Processing Systems},
 editor = {S. Koyejo and S. Mohamed and A. Agarwal and D. Belgrave and K. Cho and A. Oh},
 pages = {7641--7654},
 publisher = {Curran Associates, Inc.},
 title = {LAMP: Extracting Text from Gradients with Language Model Priors},
 volume = {35},
 year = {2022}
}

@inproceedings{deng2021tag,
  title={Tag: Gradient attack on transformer-based language models},
  author={Deng, Jieren and Wang, Yijue and Li, Ji and Wang, Chenghong and Shang, Chao and Liu, Hang and Rajasekaran, Sanguthevar and Ding, Caiwen},
  booktitle={Findings of the association for computational linguistics: EMNLP 2021},
  pages={3600--3610},
  year={2021},
  address = "Punta Cana, Dominican Republic",
  publisher = "Association for Computational Linguistics",
}

@article{chen2020practical,
  title={Practical membership inference attack against collaborative inference in industrial IoT},
  author={Chen, Hanxiao and Li, Hongwei and Dong, Guishan and Hao, Meng and Xu, Guowen and Huang, Xiaoming and Liu, Zhe},
  journal={IEEE Transactions on Industrial Informatics},
  volume={18},
  number={1},
  pages={477--487},
  year={2020},
  publisher={IEEE},
}

@article{yi2025membership,
  title={Membership inference attacks against split inference via knowledge transfer},
  author={Yi, Wenzhe and Yang, Mengda and Wang, Juan and Hu, Hongxin and Li, Ziang and Xu, Xiaoyang and He, Yu and Wang, Yao},
  journal={Neurocomputing},
  pages={132247},
  year={2025},
  publisher={Elsevier}
}

@inproceedings{mao2023secure,
  title={Secure split learning against property inference, data reconstruction, and feature space hijacking attacks},
  author={Mao, Yunlong and Xin, Zexi and Li, Zhenyu and Hong, Jue and Yang, Qingyou and Zhong, Sheng},
  booktitle={European Symposium on Research in Computer Security},
  pages={23--43},
  year={2023},
  organization={Springer}
}

@article{gu2017badnets,
  title={BadNets: Evaluating Backdooring Attacks on Deep Neural Networks},
  author={Tianyu Gu and Kang Liu and Brendan Dolan-Gavitt and Siddharth Garg},
  journal={IEEE Access},
  year={2019},
  volume={7},
  pages={47230-47244},
}

@article{yu2023backdoor,
  title={How to backdoor split learning},
  author={Yu, Fangchao and Wang, Lina and Zeng, Bo and Zhao, Kai and Pang, Zhi and Wu, Tian},
  journal={Neural Networks},
  volume={168},
  pages={326--336},
  year={2023},
  publisher={Elsevier}
}

@article{he2023backdoor,
  title={Backdoor attack against split neural network-based vertical federated learning},
  author={He, Ying and Shen, Zhili and Hua, Jingyu and Dong, Qixuan and Niu, Jiacheng and Tong, Wei and Huang, Xu and Li, Chen and Zhong, Sheng},
  journal={IEEE Transactions on Information Forensics and Security},
  volume={19},
  pages={748--763},
  year={2023},
  publisher={IEEE},
}

@inproceedings{bai2023villain,
  author = {Bai, Yijie and Chen, Yanjiao and Zhang, Hanlei and Xu, Wenyuan and Weng, Haiqin and Goodman, Dou},
  title = {VILLAIN: backdoor attacks against vertical split learning},
  year = {2023},
  isbn = {978-1-939133-37-3},
  publisher = {USENIX Association},
  booktitle = {Proceedings of the 32nd USENIX Conference on Security Symposium},
  address  = {Anaheim, CA, USA},
}

@inproceedings{tajalli2023feasibility,
  title={On feasibility of server-side backdoor attacks on split learning},
  author={Tajalli, Behrad and Ersoy, O{\u{g}}uzhan and Picek, Stjepan},
  booktitle={2023 IEEE security and privacy workshops (SPW)},
  pages={84--93},
  year={2023},
  organization={IEEE}
}

@misc{pu2024dullahan,
  title={Dullahan: Stealthy Backdoor Attack against Without-Label-Sharing Split Learning}, 
  author={Yuwen Pu and Zhuoyuan Ding and Jiahao Chen and Chunyi Zhou and Qingming Li and Chunqiang Hu and Shouling Ji},
  year={2024},
  eprint={2405.12751},
  archivePrefix={arXiv},
  url={https://arxiv.org/abs/2405.12751}, 
}

@inproceedings{yu2024chronic,
  title={Chronic poisoning: Backdoor attack against split learning},
  author={Yu, Fangchao and Zeng, Bo and Zhao, Kai and Pang, Zhi and Wang, Lina},
  booktitle={Proceedings of the AAAI Conference on Artificial Intelligence},
  volume={38},
  number={15},
  pages={16531--16538},
  year={2024}
}

@inproceedings{xi2025ibas,
  title={IBAS: Imperceptible Backdoor Attacks in Split Learning with Limited Information},
  author={Xi, Peng and Peng, Shaoliang and Tang, Wenjuan},
  booktitle={Proceedings of the AAAI Conference on Artificial Intelligence},
  volume={39},
  number={26},
  pages={27715--27722},
  year={2025}
}

@inproceedings{kohankhaki2023detecting,
  title={Detecting data poisoning in split learning using intraclass-distance inflated loss},
  author={Kohankhaki, Mohammad and Ayad, Ahmad and Barhoush, Mahdi and Schmeink, Anke},
  booktitle={2023 IEEE Globecom Workshops (GC Wkshps)},
  pages={2091--2096},
  year={2023},
  organization={IEEE},
}

@inproceedings{gajbhiye2022dpa,
  title={Data poisoning attack by label flipping on splitfed learning},
  author={Gajbhiye, Saurabh and Singh, Priyanka and Gupta, Shaifu},
  booktitle={International Conference on Recent Trends in Image Processing and Pattern Recognition},
  pages={391--405},
  year={2022},
  organization={Springer},
  publisher="Springer Nature Switzerland",
  address="Cham"
}

@article{ismail2023analyzing,
  title={Analyzing the vulnerabilities in Split Federated Learning: assessing the robustness against data poisoning attacks},
  author={Aysha-Thahsin Zahir-Ismail and Raj Mani Shukla},
  journal={Scientific Reports},
  year={2025},
  volume={15},
}

@article{wu2024evaluating,
  title={Evaluating security and robustness for split federated learning against poisoning attacks},
  author={Wu, Xiaodong and Yuan, Henry and Li, Xiangman and Ni, Jianbing and Lu, Rongxing},
  journal={IEEE Transactions on Information Forensics and Security},
  year={2024},
  publisher={IEEE}
}

@article{liu2026ppoison,
  title={PPoison: A Pluggable Poisoning attack against distributed training of split learning},
  author={Liu, Junlin and Lyu, Xinchen and Zheng, Longfei and Ren, Chenshan and Cui, Qimei},
  journal={Future Generation Computer Systems},
  volume={175},
  pages={108045},
  year={2026},
  publisher={Elsevier}
}

@inproceedings{wan2024misa,
  title={Misa: Unveiling the vulnerabilities in split federated learning},
  author={Wan, Wei and Ning, Yuxuan and Hu, Shengshan and Xue, Lulu and Li, Minghui and Zhang, Leo Yu and Jin, Hai},
  booktitle={ICASSP 2024-2024 IEEE International Conference on Acoustics, Speech and Signal Processing (ICASSP)},
  pages={6435--6439},
  year={2024},
  organization={IEEE}
}

@inproceedings{dwork2006calibrating,
  title={Calibrating noise to sensitivity in private data analysis},
  author={Dwork, Cynthia and McSherry, Frank and Nissim, Kobbi and Smith, Adam},
  booktitle={Theory of cryptography conference},
  pages={265--284},
  year={2006},
  organization={Springer},
  publisher="Springer",
  address="Berlin, Heidelberg",
}

@inproceedings{chatzikokolakis2013dxp,
  title={Broadening the scope of differential privacy using metrics},
  author={Chatzikokolakis, Konstantinos and Andr{\'e}s, Miguel E and Bordenabe, Nicol{\'a}s Emilio and Palamidessi, Catuscia},
  booktitle={international symposium on privacy enhancing technologies symposium},
  year={2013},
  publisher="Springer Berlin Heidelberg",
  address="Berlin, Heidelberg",
}

@inproceedings{yue2021santext,
  title={Differential privacy for text analytics via natural text sanitization},
  author={Yue, Xiang and Du, Minxin and Wang, Tianhao and Li, Yaliang and Sun, Huan and Chow, Sherman SM},
  booktitle={Findings of the Association for Computational Linguistics: ACL-IJCNLP 2021},
  pages={3853--3866},
  year={2021}
}

@inproceedings{chen2023custext,
  title={A customized text sanitization mechanism with differential privacy},
  author={Chen, Sai and Mo, Fengran and Wang, Yanhao and Chen, Cen and Nie, Jian-Yun and Wang, Chengyu and Cui, Jamie},
  booktitle={Findings of the Association for Computational Linguistics},
  pages={5747--5758},
  year={2023},
  address = "Toronto, Canada",
}

@article{tong2025inferdpt,
  title={Inferdpt: Privacy-preserving inference for black-box large language models},
  author={Tong, Meng and Chen, Kejiang and Zhang, Jie and Qi, Yuang and Zhang, Weiming and Yu, Nenghai and Zhang, Tianwei and Zhang, Zhikun},
  journal={IEEE Transactions on Dependable and Secure Computing},
  year={2025},
  publisher={IEEE}
}

@inproceedings{wu2025cape,
  title={Cape: Context-Aware Prompt Perturbation Mechanism with Differential Privacy},
  author={Wu, Haoqi and Dai, Wei and Li, Wang and Yan, Qiang},
  booktitle={International Conference on Machine Learning},
  pages={67184--67201},
  year={2025},
  organization={PMLR}
}

@inproceedings{du2023dpforward,
  title={Dp-forward: Fine-tuning and inference on language models with differential privacy in forward pass},
  author={Du, Minxin and Yue, Xiang and Chow, Sherman SM and Wang, Tianhao and Huang, Chenyu and Sun, Huan},
  booktitle={Proceedings of the 2023 ACM SIGSAC Conference on Computer and Communications Security},
  pages={2665--2679},
  year={2023},
  address = {New York, NY, USA}, 
}

@inproceedings{pan2020privacy,
  title={Privacy risks of general-purpose language models},
  author={Pan, Xudong and Zhang, Mi and Ji, Shouling and Yang, Min},
  booktitle={2020 IEEE Symposium on Security and Privacy (SP)},
  pages={1314--1331},
  year={2020},
  organization={IEEE}
}

@article{wu2023federated,
  title={Federated split learning with data and label privacy preservation in vehicular networks},
  author={Wu, Maoqiang and Cheng, Guoliang and Ye, Dongdong and Kang, Jiawen and Yu, Rong and Wu, Yuan and Pan, Miao},
  journal={IEEE Transactions on Vehicular Technology},
  volume={73},
  number={1},
  pages={1223--1238},
  year={2023},
  publisher={IEEE}
}

@misc{yang2022differentially,
  title={Differentially Private Label Protection in Split Learning}, 
  author={Xin Yang and Jiankai Sun and Yuanshun Yao and Junyuan Xie and Chong Wang},
  year={2022},
  eprint={2203.02073},
  archivePrefix={arXiv},
  url={https://arxiv.org/abs/2203.02073}, 
}

@misc{qiu2023evaluating,
  title={Evaluating Privacy Leakage in Split Learning}, 
  author={Xinchi Qiu and Ilias Leontiadis and Luca Melis and Alex Sablayrolles and Pierre Stock},
  year={2024},
  eprint={2305.12997},
  archivePrefix={arXiv},
  url={https://arxiv.org/abs/2305.12997}, 
}

@inproceedings{shen2025sap,
  title={SAP: Privacy-Preserving Fine-Tuning on Language Models with Split-and-Privatize Framework},
  author={Shen, Xicong and Liu, Yang and Liu, Yi and Wang, Peiran and Liu, Huiqi and Hong, Jue and Duan, Bing and Huang, Zirui and Mao, Yunlong and Wu, Ye and others},
  booktitle={Proceedings of the Thirty-Fourth International Joint Conference on Artificial Intelligence},
  pages={502--510},
  year={2025}
}

@article{wang2024selective,
  title={Selective privacy-preserving framework for large language models fine-tuning},
  author={Wang, Teng and Zhai, Lindong and Yang, Tengfei and Luo, Zhucheng and Liu, Shuanggen},
  journal={Information Sciences},
  volume={678},
  pages={121000},
  year={2024},
  publisher={Elsevier}
}

@inproceedings{mai2023snd,
  title={Split-and-denoise: protect large language model inference with local differential privacy},
  author={Mai, Peihua and Yan, Ran and Huang, Zhe and Yang, Youjia and Pang, Yan},
  booktitle={Proceedings of the 41st International Conference on Machine Learning},
  pages={34281--34302},
  year={2024}
}

@misc{liu2024mitigating,
  title={Mitigating privacy risks in LLM embeddings from embedding inversion},
  author={Liu, Tiantian and Yao, Hongwei and Wu, Tong and Qin, Zhan and Lin, Feng and Ren, Kui and Chen, Chun},
  eprint={2511.23167},
  archivePrefix={arXiv},
  year={2024},
  url={https://arxiv.org/html/2411.05034v1}
}

@misc{zou2023mutual,
  title={Mutual Information Regularization for Vertical Federated Learning}, 
  author={Tianyuan Zou and Yang Liu and Ya-Qin Zhang},
  year={2023},
  eprint={2301.01142},
  archivePrefix={arXiv},
  url={https://arxiv.org/abs/2301.01142}, 
}

@inproceedings{zhou2023textobfuscator,
  title={TextObfuscator: Making pre-trained language model a privacy protector via obfuscating word representations},
  author={Zhou, Xin and Lu, Yi and Ma, Ruotian and Gui, Tao and Wang, Yuran and Ding, Yong and Zhang, Yibo and Zhang, Qi and Huang, Xuan-Jing},
  booktitle={Findings of the association for computational linguistics: ACL 2023},
  pages={5459--5473},
  year={2023},
  address = "Toronto, Canada",
}

@inproceedings{wang2021improving,
  title={Improving robustness to model inversion attacks via mutual information regularization},
  author={Wang, Tianhao and Zhang, Yuheng and Jia, Ruoxi},
  booktitle={Proceedings of the AAAI conference on artificial intelligence},
  volume={35},
  number={13},
  pages={11666--11673},
  year={2021}
}

@inproceedings{vepakomma2020nopeek,
  title={NoPeek: Information leakage reduction to share activations in distributed deep learning},
  author={Vepakomma, Praneeth and Singh, Abhishek and Gupta, Otkrist and Raskar, Ramesh},
  booktitle={2020 International Conference on Data Mining Workshops (ICDMW)},
  pages={933--942},
  year={2020},
  organization={IEEE}
}

@article{pham2024enhancing,
  title={Enhancing accuracy-privacy trade-off in differentially private split learning},
  author={Pham, Ngoc Duy and Phan, Khoa T and Chilamkurti, Naveen},
  journal={IEEE Transactions on Emerging Topics in Computational Intelligence},
  volume={9},
  number={1},
  pages={988--1000},
  year={2024},
  publisher={IEEE}
}

@inproceedings{erdougan2024splitout,
  title={SplitOut: Out-of-the-box training-hijacking detection in split learning via outlier detection},
  author={Erdo{\u{g}}an, Ege and Tek{\c{s}}en, Unat and {\c{C}}elikteny{\i}ld{\i}z, M Salih and K{\"u}p{\c{c}}{\"u}, Alptekin and {\c{C}}i{\c{c}}ek, A Erc{\"u}ment},
  booktitle={International Conference on Cryptology and Network Security},
  pages={118--142},
  year={2024},
  publisher = "Springer",
  address="Singapore",
}

@inproceedings{rieger2025safesplit,
  title={SafeSplit: A Novel Defense Against Client-Side Backdoor Attacks in Split Learning},
  author={Rieger, Phillip and Pegoraro, Alessandro and Kumari, Kavita and Abera, Tigist and Knauer, Jonathan and Sadeghi, Ahmad-Reza},
  booktitle = {Proceedings of the Network and Distributed System Security Symposium},
  year={2025},
  address={Reston, VA},
}

@article{brown2020language,
  title={Language models are few-shot learners},
  author={Brown, Tom and Mann, Benjamin and Ryder, Nick and Subbiah, Melanie and Kaplan, Jared D and Dhariwal, Prafulla and Neelakantan, Arvind and Shyam, Pranav and Sastry, Girish and Askell, Amanda and others},
  journal={Advances in neural information processing systems},
  volume={33},
  pages={1877--1901},
  address = {Vancouver, BC, Canada},
  publisher = {Curran Associates Inc.},
  year={2020}
}

@article{radford2018improving,
  title={Improving language understanding by generative pre-training},
  author={Radford, Alec and Narasimhan, Karthik and Salimans, Tim and Sutskever, Ilya and others},
  year={2018},
  publisher={San Francisco, CA, USA}
}

@misc{yang2025qwen3,
  title={Qwen3 technical report},
  author={Yang, An and Li, Anfeng and Yang, Baosong and Zhang, Beichen and Hui, Binyuan and Zheng, Bo and Yu, Bowen and Gao, Chang and Huang, Chengen and Lv, Chenxu and others},
  year={2025},
  eprint={2505.09388},
  archivePrefix={arXiv},
  url={https://arxiv.org/abs/2505.09388}, 
}

@inproceedings{ayad2021improving,
  title={Improving the communication and computation efficiency of split learning for iot applications},
  author={Ayad, Ahmad and Renner, Melvin and Schmeink, Anke},
  pages={01--06},
  year={2021},
  booktitle = {Proceedings of the 2021 IEEE Global Communications Conference},
  location = {Madrid, Spain},
  pages = {01--06},
  numpages = {6},
  publisher = {IEEE},
  address = {Piscataway, NJ, USA},
}

@article{liu2024deepseek,
  title={Deepseek-v3 technical report},
  author={Liu, Aixin and Feng, Bei and Xue, Bing and Wang, Bingxuan and Wu, Bochao and Lu, Chengda and Zhao, Chenggang and Deng, Chengqi and Zhang, Chenyu and Ruan, Chong and others},
  journal={arXiv preprint arXiv:2412.19437},
  year={2025},
  eprint={2412.19437},
  archivePrefix={arXiv},
  url={https://arxiv.org/abs/2412.19437}, 
}

@inproceedings{li2022ressfl,
  title={Ressfl: A resistance transfer framework for defending model inversion attack in split federated learning},
  author={Li, Jingtao and Rakin, Adnan Siraj and Chen, Xing and He, Zhezhi and Fan, Deliang and Chakrabarti, Chaitali},
  booktitle={Proceedings of the IEEE/CVF conference on computer vision and pattern recognition},
  pages={10194--10202},
  year={2022},
}

@article{Liu2025GSFLAP,
  title={GSFL: A Privacy-Preserving Grouping-Split Federated Learning Approach in Resource-Constrained Edge Computing Scenarios},
  author={Qi Liu and Zhilu Wang and Xiaokang Zhou and Yonghong Zhang and Xiaodong Liu and Haiyang Lin},
  journal={ACM Transactions on Autonomous and Adaptive Systems},
  year={2025},
  volume={20},
  pages={1 - 29},
}

@inproceedings{zhang2023cluster,
  title={Cluster-hsfl: a cluster-based hybrid split and federated learning},
  author={Zhang, Songge and Tu, Haoyu and Li, Zuguang and Liu, Shengbo and Li, Shaofeng and Wu, Wen and Shen, Xuemin Sherman},
  booktitle={2023 IEEE/CIC International Conference on Communications in China (ICCC)},
  pages={1--2},
  year={2023},
  organization={IEEE}
}

@inproceedings{xu2024csfl,
  title={CSFL: Enhancing Splitfed Learning with Clustering on Non-IID Data},
  author={Xu, Guangwei and Lu, Jianbo and Wang, Xinru and Lu, Yang and Cao, Mei and Zhao, Mengying},
  booktitle={2024 IEEE International Conference on High Performance Computing and Communications (HPCC)},
  pages={791--798},
  year={2024},
  organization={IEEE}
}

@article{deng2026clicooper,
  title={Client-Cooperative Split Learning},
  author={Deng, Haiyu and Jiang, Yanna and Yu, Guangsheng and Wang, Qin and Wang, Xu and Ni, Wei and Chen, Shiping and Liu, Ren Ping},
  journal={IEEE Transactions on Services Computing},
  year={2026},
  publisher={IEEE},
}

@inproceedings{shi2025navigating,
  title={Navigating the Designs of Privacy-Preserving Fine-tuning for Large Language Models},
  author={Shi, Haonan and Ouyang, Tu and Wang, An},
  booktitle={Companion Proceedings of the ACM on Web Conference 2025},
  pages={1298--1302},
  year={2025},
  address = {New York, NY, USA},
}

@article{chen2025pasfl,
  title={Privacy-aware split federated learning for LLM fine-tuning over internet of things},
  author={Chen, Xiaopei and Wu, Wen and Ji, Fei and Lu, Yongguang and Li, Liang},
  journal={IEEE Internet of Things Journal},
  year={2025},
  publisher={IEEE},
}

@article{fu2024projpert,
  title={ProjPert: Projection-based perturbation for label protection in split learning based vertical federated learning},
  author={Fu, Fangcheng and Wang, Xuanyu and Jiang, Jiawei and Xue, Huanran and Cui, Bui},
  journal={IEEE Transactions on Knowledge and Data Engineering},
  volume={36},
  number={7},
  pages={3417--3428},
  year={2024},
  publisher={IEEE},
}

@preprint{gu2026del,
  title={Differentially Private and Communication Efficient Large Language Model Split Inference via Stochastic Quantization and Soft Prompt},
  author={Gu, Yujie and Jin, Richeng and Ji, Xiaoyu and Jin, Yier and Xu, Wenyuan},
  eprint={2602.11513},
  archivePrefix={arXiv},
  year={2026},
  url={https://arxiv.org/abs/2602.11513}, 
}

@article{xia2025sfml,
  title={SFML: A personalized, efficient, and privacy-preserving collaborative traffic classification architecture based on split learning and mutual learning},
  author={Xia, Jiaqi and Wu, Meng and Li, Pengyong},
  journal={Future Generation Computer Systems},
  volume={162},
  pages={107487},
  year={2025},
  publisher={Elsevier}
}

@article{huang2025sldp,
  title={SLDP-LoRA: A privacy-preserving split learning framework with low-rank adaptation},
  author={Huang, Yirui and Yin, Jia-Li and Tan, Zhou and Wang, Qiuxiang and Liu, Ximeng},
  journal={IEEE Transactions on Network Science and Engineering},
  year={2026},
  publisher={IEEE},
  pages={2111-2127},
  volume={13},
}

@article{pham2023binarizing,
  title={Binarizing split learning for data privacy enhancement and computation reduction},
  author={Pham, Ngoc Duy and Abuadbba, Alsharif and Gao, Yansong and Phan, Khoa Tran and Chilamkurti, Naveen},
  journal={IEEE Transactions on Information Forensics and Security},
  volume={18},
  pages={3088--3100},
  year={2023},
  publisher={IEEE}
}

@misc{kanpak2024cure,
  title={CURE: Privacy-Preserving Split Learning Done Right}, 
  author={Halil Ibrahim Kanpak and Aqsa Shabbir and Esra Genç and Alptekin Küpçü and Sinem Sav},
  year={2024},
  eprint={2407.08977},
  archivePrefix={arXiv},
  url={https://arxiv.org/abs/2407.08977}, 
}

@misc{pereteanu2022splithe,
  title={Split HE: Fast Secure Inference Combining Split Learning and Homomorphic Encryption}, 
  author={George-Liviu Pereteanu and Amir Alansary and Jonathan Passerat-Palmbach},
  year={2022},
  eprint={2202.13351},
  archivePrefix={arXiv},
  url={https://arxiv.org/abs/2202.13351}, 
}

@article{liang2025dahe,
  title={Federated split learning via dynamic aggregation and homomorphic encryption on non-IID data},
  author={Liang, Xingzhu and Xu, Yachen and Lin, Yu-e and Zhang, Chunjiong},
  journal={The Journal of Supercomputing},
  volume={81},
  number={1},
  pages={63},
  year={2025},
  publisher={Springer},
}

@article{hou2026ciphergpt,
  title={Ciphergpt: Secure two-party gpt inference},
  author={Hou, Xiaoyang and Liu, Jian and Li, Jingyu and Li, Yuhan and Zhang, Jiawen and Lu, Wen-jie and Hong, Cheng and Ren, Kui},
  journal={IEEE Transactions on Dependable and Secure Computing},
  year={2026},
  publisher={IEEE},
  pages={1-16},
}

@inproceedings{chen2022thex,
  title={The-x: Privacy-preserving transformer inference with homomorphic encryption},
  author={Chen, Tianyu and Bao, Hangbo and Huang, Shaohan and Dong, Li and Jiao, Binxing and Jiang, Daxin and Zhou, Haoyi and Li, Jianxin and Wei, Furu},
  booktitle={Findings of the association for computational linguistics: ACL 2022},
  pages={3510--3520},
  year={2022},
  address = "Dublin, Ireland",
  publisher = "Association for Computational Linguistics",
}

@article{lu2023bumblebee,
  title={Bumblebee: Secure two-party inference framework for large transformers},
  author={Lu, Wen-jie and Huang, Zhicong and Gu, Zhen and Li, Jingyu and Liu, Jian and Hong, Cheng and Ren, Kui and Wei, Tao and Chen, WenGuang},
  journal={Cryptology ePrint Archive},
  year={2023}
}

@inproceedings{alromih2022privacy,
author = {Alromih, Arwa and Clark, John A. and Gope, Prosanta},
title = {Privacy-Aware Split Learning Based Energy Theft Detection for Smart Grids},
year = {2022},
isbn = {978-3-031-15776-9},
publisher = {Springer-Verlag},
address = {Berlin, Heidelberg},
booktitle = {Information and Communications Security: 24th International Conference, ICICS 2022, Canterbury, UK, September 5–8, 2022, Proceedings},
pages = {281–300},
numpages = {20},
location = {Canterbury, United Kingdom}
}

@article{pham2025split,
  title={Split learning without local weight sharing to enhance client-side data privacy},
  author={Pham, Ngoc Duy and Phan, Tran Khoa and Abuadbba, Alsharif and Gao, Yansong and Nguyen, Van-Doan and Chilamkurti, Naveen},
  journal={IEEE Transactions on Dependable and Secure Computing},
  year={2025},
  publisher={IEEE}
}

@inproceedings{jiang2025splitunlearning,
  title={Split Unlearning},
  author={Jiang, Yanna and Yu, Guangsheng and Wang, Qin and Wang, Xu and Ma, Baihe and Sun, Caijun and Ni, Wei and Liu, Ren Ping},
  booktitle={Proceedings of the 2025 ACM SIGSAC Conference on Computer and Communications Security},
  pages={948--962},
  numpages = {15},
  year={2025},
  publisher = {Association for Computing Machinery},
  address = {New York, NY, USA},
}

@inproceedings{bourtoule2021machine,
  title={Machine unlearning},
  author={Bourtoule, Lucas and Chandrasekaran, Varun and Choquette-Choo, Christopher A and Jia, Hengrui and Travers, Adelin and Zhang, Baiwu and Lie, David and Papernot, Nicolas},
  booktitle={2021 IEEE symposium on security and privacy (SP)},
  pages={141--159},
  year={2021},
  organization={IEEE},
}

@misc{minaee2024llmsurvey,
  title={Large Language Models: A Survey}, 
  author={Shervin Minaee and Tomas Mikolov and Narjes Nikzad and Meysam Chenaghlu and Richard Socher and Xavier Amatriain and Jianfeng Gao},
  year={2025},
  eprint={2402.06196},
  archivePrefix={arXiv},
  url={https://arxiv.org/abs/2402.06196}, 
}

@inproceedings{liang2025towards,
  title={Towards Straggler-Resilient Split Federated Learning: An Unbalanced Update Approach},
  author={Liang, Dandan and Zhang, Jianing and Chen, Evan and Li, Zhe and Li, Rui and Yang, Haibo},
  booktitle={The Thirty-ninth Annual Conference on Neural Information Processing Systems},
  year={2025},
  address = {Red Hook, NY, USA},
}


\end{document}